# A Comprehensive Review of Heat Transfer in Thermoelectric Materials and Devices

Zhiting Tian, Sangyeop Lee and Gang Chen

*Department of Mechanical Engineering*

*Massachusetts Institute of Technology*

*Cambridge, MA 02139, USA*

**Abstract**

Solid-state thermoelectric devices are currently used in applications ranging from thermocouple sensors to power generators in satellites, to portable air-conditioners and refrigerators. With the ever-rising demand throughout the world for energy consumption and $CO_2$ reduction, thermoelectric energy conversion has been receiving intensified attention as a potential candidate for waste-heat harvesting as well as for power generation from renewable sources. Efficient thermoelectric energy conversion critically depends on the performance of thermoelectric materials and devices. In this review, we discuss heat transfer in thermoelectric materials and devices, especially phonon engineering to reduce the lattice thermal conductivity of thermoelectric materials, which requires a fundamental understanding of nanoscale heat conduction physics.

**Nomenclature**
Constant:
$k_B$ = Boltzmann constant, $1.38 \times 10^{-23} m^2 kg s^{-2}$
$\hbar$ = reduced Planck constant, $1.055 \times 10^{-34} Js$

Symbols:
A = cross-sectional area, $m^2$
C = spectral volumetric specific heat, J $m^{-3}$ $Hz^{-1}$ $K^{-1}$
D = density of states per unit volume per unit frequency interval, $m^{-3}$ $Hz^{-1}$
$e$ = charge per carrier, C
$E_f$ = Fermi level (i.e. electrochemical potential) relative to conduction band edge $E_c$, J
$E_g$ = band gap energy
$f_0$ = Fermi-Dirac distribution
$f_{BE}$ = Bose-Einstein distribution
I = electrical current, A
$J_e$ = electrical current density, A $m^{-2}$
$J_q$ = heat flux, W $m^{-2}$
K = thermal conductance, W $m^{-2}$ $K^{-1}$

*L* = transport coefficients
$\mathcal{L}$ = Lorenz number, W Ω K$^{-2}$
*n* = carrier concentration m$^{-3}$
*Q* = heat current, W
*R* = electrical resistance, Ω
*S* = Seebeck coefficient, V K$^{-1}$
*T* = temperature, K
*v* = velocity, m s$^{-1}$
*Z* = figure of merit, K$^{-1}$
*B* = Thomson coefficient, V K$^{-1}$
Π = Peltier coefficient, V
η = efficiency
κ = thermal conductivity, W m$^{-1}$ K$^{-1}$
Φ = electrochemical potential, V
ϕ = coefficient of performance
*ρ* = electrical resistivity, Ω m
σ = electrical conductivity, Ω$^{-1}$ m$^{-1}$
τ = lifetime, s
ω = angular frequency, rad·Hz

Subscript:
C: cold side
H: hot side
i: inlet
M: mean
n: n-type
o: outlet
p: p-type

Abbreviations:
AMM = acoustic mismatch model
BTE = Boltzmann transport equation
COP = coefficient of performance
DFT = density functional theory
DMM = diffuse mismatch model
EMA = effective medium approach
EMD = equilibrium molecular dynamics
MC = Monte Carlo
MD = molecular dynamics
MFP = mean free path
NEMD = non-equilibrium molecular dynamics

## 1. Introduction

Thermoelectric effects have long been known since the Seebeck effect and the Peltier effect were discovered in 1800s [1]. The Seebeck effect describes the phenomenon that a voltage is generated in a conductor or semiconductor subjected to a temperature gradient. This effect is the basis of thermocouples and can be applied to thermal to electrical energy conversion. The inverse process, in which an electrical current creates cooling or heat pumping at the junction between two dissimilar materials, is called the Peltier effect. The decade of 1950s saw extensive research and applications on Peltier refrigerators following the emergence of semiconductors and their alloys as thermoelectric materials [1]. However, the research efforts waned as the efficiency of solid-state refrigerators could not compete with mechanical compression cycles. Starting in 1990s, interest in thermoelectrics renewed because of increased global energy demand and global warming caused by excessive $CO_2$ emissions [2,3]. The maximum efficiency of a thermoelectric device for both thermoelectric power generation and cooling is determined by the dimensionless figure-of-merit, ZT,

$$ZT = \frac{S^2 \sigma}{\kappa} T \qquad (1.1)$$

where $S$ is the Seebeck coefficient, $\sigma$ is the electrical conductivity, $S^2\sigma$ is the power factor, and $\kappa = \kappa_p + \kappa_e$ is the thermal conductivity which is composed of lattice (phononic) thermal conductivity $\kappa_p$ and electronic thermal conductivity $\kappa_e$. The Seebeck coefficient is voltage generated per degree of temperature difference over a material. Fundamentally, it is a measure of the average entropy carried by a charge in the material. A large power factor means electrons are efficient in the heat-electricity conversion, while a small thermal conductivity is required to maintain a temperature gradient and reduce conduction heat losses. Achieving high ZT values requires a material simultaneously possessing a high Seebeck coefficient and a high electrical conductivity, but maintaining a low thermal conductivity, which is challenging as these requirements are often contradictory to each other in conventional materials [4].

Recent years have witnessed impressive progresses in thermoelectric materials. There have been many reviews [3-24] emphasizing different aspects of thermoelectrics, including bulk thermoelectric materials [10,12], individual nanostructures [6,13,15,25], bulk nanostructures [7,8], and interfaces in bulk thermoelectric materials [16]. In this paper, we intend to emphasize the heat transfer aspects of thermoelectric research at both materials and system levels. We will start with a brief summary of the basic principles of thermoelectric devices and materials (Sec. 2), followed by a summary of materials status (Sec. 3). We will then discuss heat conduction mechanisms in bulk materials (Sec.4), and strategies to reduce thermal conductivity in bulk materials (Sec.5) and using nanostructures (Sec.6). Heat transfer issues at device and system levels will be presented in Sec. 7. A shorter version of this paper will appear in the Journal of Heat Transfer due to page limit. This review is more comprehensive, including an introductory section on thermoelectric transport and device basics (Sec. 2), more detailed discussion on heat transfer in thermoelectric devices and systems (Sec. 7) and 16 more figures.

**2. Thermoelectric Transport and Device Basics**

## 2.1 Thermoelectric Effects and Transport Properties

Thermoelectric effects dictate the coupling between heat and electricity. Fundamentally, charges in semiconductors and conductors carry heat with them. The coupled charge current flux, $J_e$, and heat current flux, $J_q$, can be expressed as [26,27]

$$J_e = L_{11}(-\frac{d\Phi}{dx}) + L_{12}(-\frac{dT}{dx}) \tag{2.1}$$

$$J_q = L_{21}(-\frac{d\Phi}{dx}) + L_{22}(-\frac{dT}{dx}) \tag{2.2}$$

for one-dimensional transport, where $L_{11}$ is the electrical conductivity and $\Phi$ is the electromotive force, i.e., the electrochemical potential divided by unit charge. In an isothermal conductor, the heat current,

$$J_q = \frac{L_{21}}{L_{11}} J_e = \Pi J_e \tag{2.3}$$

is proportional to the electrical current and the proportionality constant is the Peltier coefficient, $\Pi$. Although the Peltier coefficient is an intrinsic material property, Peltier cooling and heating happens when two materials with different Peltier coefficients are joined together, as shown in Fig. 1, due to the imbalance of the Peltier heat flowing in and out of the junction. A control volume analysis shows that the cooling or heating, Q, happening at the junction is equal to

$$Q = (\Pi_2 - \Pi_1)I \tag{2.4}$$

where I is the electrical current, and subscripts 1 and 2 represent Peltier coefficients for the two materials. Under open voltage condition, Eq. (2.1) leads to the Seebeck coefficient

$$S = \frac{-d\Phi/dx}{dT/dx} = -\frac{V}{\Delta T} = \frac{L_{12}}{L_{11}} \tag{2.5}$$

From the definition of the Seebeck coefficient, we can see that the Seebeck voltage *V* between two points on a homogenous material does not depend on the temperature profile. This fact is used widely in temperature measurements by thermocouples.

The Seebeck and Peltier coefficients are related to each other through, $\Pi = TS$, which is an example of the generalized Onsager reciprocity relation $L_{21}=TL_{12}$.

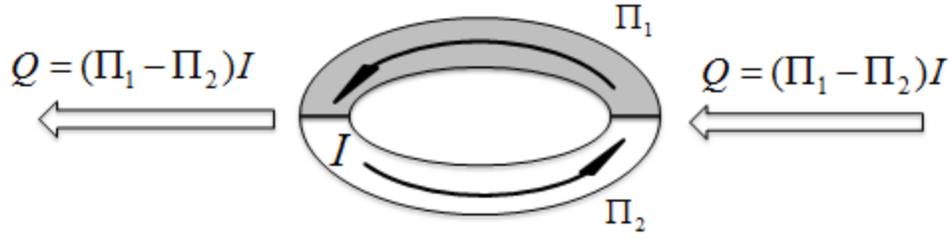

Fig. 1 Cooling or heating at the junctions of two materials occur because of the different Peltier coefficients of the two materials, which can be used as thermoelectric refrigerators or heat pumps

Furthermore, one can eliminate the electrochemical potential in Eq. (2.1) and express the heat current as

$$J_q = \Pi J_e - \kappa_e \frac{dT}{dx} \qquad (2.6)$$

where the electronic contribution to thermal conductivity is

$$\kappa_e = L_{22} - \frac{L_{12}L_{21}}{L_{11}} \qquad (2.7)$$

The above expression has not included the phonon contribution to thermal conductivity. However, as long as one can separate the electronic and phononic components, one can simply add an additional phononic term to Eq. (2.6) for heat flowing in a solid.

Equation (2.6) can be used with the first law of thermodynamics, leading to the following equation that determines the temperature distribution in a thermoelectric material [28]

$$C_p \frac{dT}{dt} = \frac{d}{dx}\left(\kappa \frac{dT}{dx}\right) - \left(T \frac{dS}{dT}\right) J_e \frac{dT}{dx} + \frac{J_e^2}{\sigma} \qquad (2.8)$$

where $C_p$ is the specific heat per unit volume at constant pressure.

The above equation contains an extra term compared to the ordinary heat conduction equation. This is due to the Thomson effect [1], which suggests that distributed heating or cooling can occur even in the same solid because of the temperature dependence of Seebeck coefficient. The Thomson coefficient is defined by

$$\beta = \frac{\dot{q}_c}{J_e \frac{dT}{dx}} = T \frac{dS}{dT} \qquad (2.9)$$

where $\dot{q}$ is the rate of cooling flux or heating per unit volume.

All three thermoelectric effects (Seebeck effect, Peltier effect, and Thomson effect) are intrinsically connected as they are simply different manifestations of the heat carried by charges. The Thomson effect is often neglected, although careful device simulation should take it into consideration. It can be seen from Eq. (2.8) that the thermoelectric equation does not include the Peltier or the Seebeck coefficient directly. Often, they form part of the boundary equations.

Transport coefficients can be further related to the electronic band structure of materials. Under the relaxation time approximation, the transport coefficients for an isotropic material with single band can be expressed as [29]

$$L_{11} = -\frac{e^2}{3} \int v^2 \tau \frac{\partial f_0}{\partial E} D(E) dE \qquad (2.10)$$

$$L_{12} = \frac{e}{3T} \int v^2 \tau (E - E_f) \frac{\partial f_0}{\partial E} D(E) dE \qquad (2.11)$$

$$L_{22} = -\frac{1}{3T} \int v^2 \tau (E - E_f)^2 \frac{\partial f_0}{\partial E} D(E) dE \qquad (2.12)$$

where $D(E)$ is the electron density of states per unit volume per unit energy interval, $f_0$ is the Fermi-Dirac distribution, and $E_f$ is the electrochemical potential that can be controlled by doping and measured from the conduction band edge $E_c$. Based on the above expression, the Seebeck coefficient can be expressed as

$$S = -\frac{1}{eT} \frac{\int v^2 \tau (E - E_f) \frac{\partial f_0}{\partial E} D(E) dE}{\int v^2 \tau \frac{\partial f_0}{\partial E} D(E) dE} \qquad (2.13)$$

which shows that the Seebeck coefficient is an average energy of an electron within the Fermi window under the open circuit condition

Taking a simple parabolic band, and assuming an energy-independent constant relaxation time, we can write the transport coefficients as

$$L_{11} = \frac{e^2}{T} K_0 \qquad (2.14)$$

$$L_{12} = \frac{e}{T^2} (E_f K_0 - K_1) \qquad (2.15)$$

$$L_{22} = \frac{1}{T^2} (E_f^2 K_0 - 2E_f K_1 + K_2) \qquad (2.16)$$

where $K_n = 2(3\pi^2\hbar^3)^{-1}(2m^*)^{1/2} T\tau(n+3/2)(k_B T)^{n+3/2} F_{n+1/2}(\eta)$ and $F_j(\eta) = \int_0^\infty E^j (e^{E-\eta}+1)^{-1} dE$ is the Fermi-Dirac integral with $\eta = E_f/(k_B T)$.

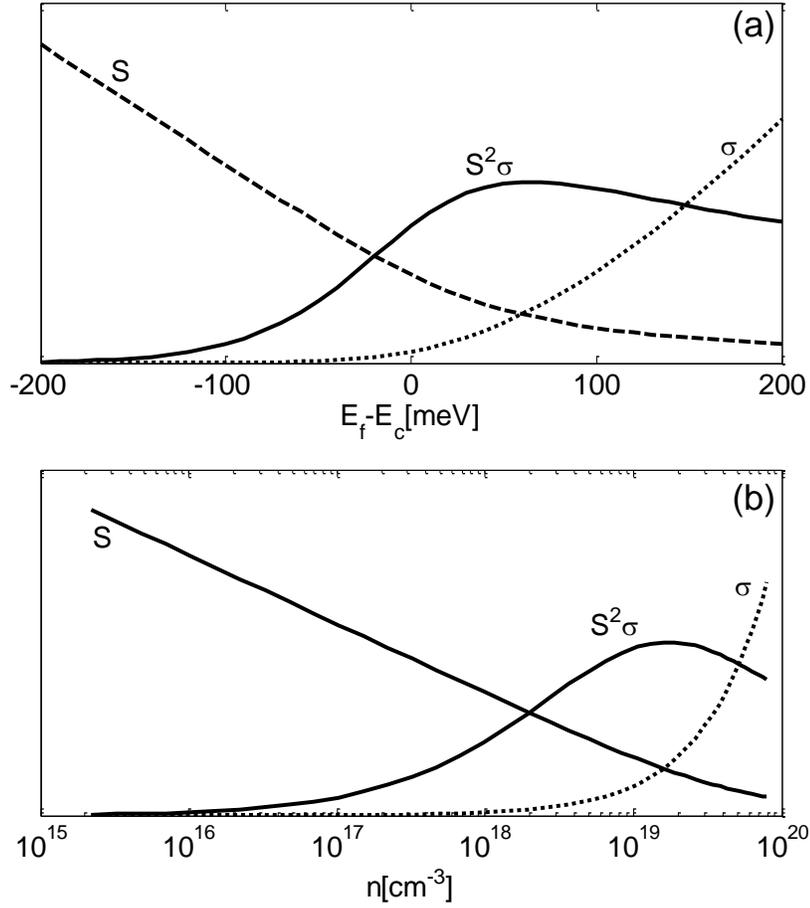

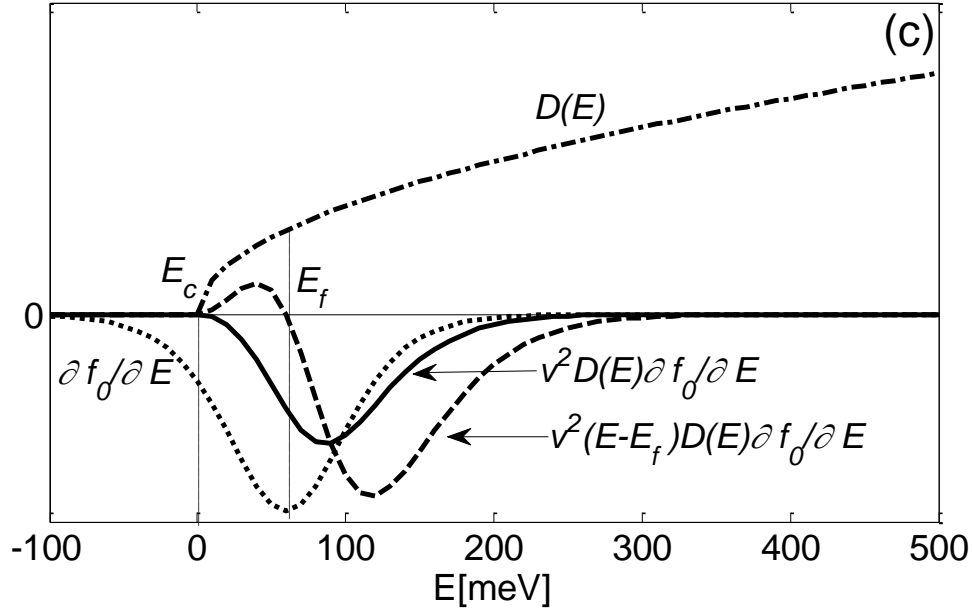

Fig. 2 Sketches of Seebeck coefficient, S, electrical conductivity, σ, and power factor, $S^2\sigma$, as a function of (a) $E_f$-$E_c$, and (b) carrier concentration, n. (c) Electron density of states, D(E), $\partial f_0/\partial E$, $v^2 D(\partial f_0/\partial E)$ and $v^2(E-E_f)D(\partial f_0/\partial E)$ versus electron energy, E. The plots are based on single parabolic band and constant relaxation time, assuming $m^* = 0.33m$ and $\tau = 10^{-12} s$. The full scale for S, σ, and $S^2\sigma$ in (a) and (b) are 0~$10^{-3}$ V/K, 0~$10^7$ $\Omega^{-1} m^{-1}$ and 0~0.05 J/m$^2$K, respectively.

In Fig. 2(a), we plot S, σ and $S^2\sigma$ as a function of $E_f$ relative to the band edge $E_c$, and in Fig.2(b), the same quantities are replotted as a function of carrier concentration calculated based on $E_f$-$E_c$. These figures show that there is an optimal value of $E_f$, i.e. carrier concentration, which maximize $S^2\sigma$. This means that thermoelectric materials require doping optimization. At the optimal $E_f$, the dopant concentration is usually large, in the order of $10^{19}$ cm$^{-3}$ as shown in Fig. 2(b) comparing to $10^{17}$-$10^{18}$ cm$^{-3}$ in electronics. In Fig.2(c), key integrand variables, the density of states D(E) and the derivative of the carrier concentration $\partial f_0/\partial E$, are plotted at the optimal carrier concentration. The products of $v^2 D(\partial f_0/\partial E)$, shown in the same figure, represents the differential electrical conductivity (normalized to the relaxation time), which shows that only a small amount of carriers near $E_f$ contribute to electrical conductivity because $\partial f_0/\partial E$ is nonzero only ~ $3k_B T$ around $E_f$. Also shown in the same figure is the integrand in the numerator of the Seebeck coefficient $v^2(E-E_f)D(\partial f_0/\partial E)$. This shows that carriers with energy below $E_f$ carry positive entropy while those above carry negative entropy, although all of them contribute to the electrical conduction. This is why metals have small Seebeck coefficients, because in metals, $E_f$ is deep inside the band, and electron distributions above and below $E_f$ are nearly symmetric, canceling out the heat they carry with them. Although these plots do not consider energy dependence of the relaxation time, clearly, a large $\tau$ (i.e., high mobility) is desired to reach large $S^2\sigma$ and the dependence of the relaxation time on carrier energy can benefit or harm the power factor.

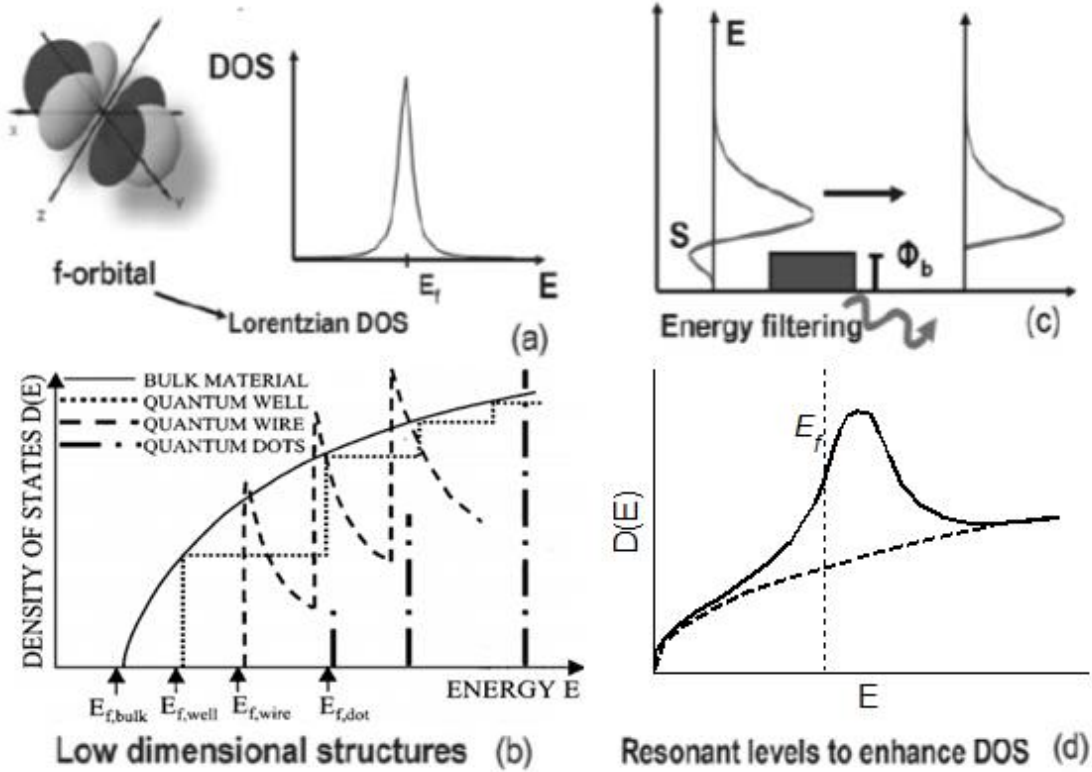

Fig. 3 Electron engineering techniques: (a) Sharp feature of the DOS in bulk materials; (b) Sharp feature of the DOS in low dimensional structures [19]; (c) Energy filtering to remove the cancellation of the Seebeck coefficient; (d) Resonant levels to enhance the DOS

For a parabolic band as shown in Fig. 2(c), typical for most bulk semiconductors, the slow rise of the density of states near $E_c$ means there are only a small number of carriers in this energy range. One must position the Fermi level deeper into the bands to increase the electrical conductivity, yet, as $E_f$ moves deeper into the bands, carriers below $E_f$ cancel the entropy carried by carriers above $E_f$. Hicks and Dresselhaus [30,31] proposed utilizing the sharp density of states in quatum structures to enhance the power factor, $S^2\sigma$. Mahan and Sofo [11] further showed that the ideal electronic density of states to maximize thermoelectric properties was a delta function, although subsequent studies showed that there was an optical bandwidth [32,33]. Mahan also identified d- and f- bands as having sharp features in bulk materials [11]. These studies provided strong stimulus for thermoelectric materials research, and many ideas in electronic band structure engineering have been proposed. Figure 3 summarizes graphically these ideas. The sharp density of states can be found in some bulk materials containing atoms with d- and f- orbitals [Fig.3(a)] or different low-dimensional systems from 1D quantum wires to 2D quantum wells and superlattices, and even 0D made of quantum dots [Fig.3(b)]. Resonant energy levels as represented by Tl-doped PbTe have sharp features in the density-of-states inside the conducting band and have been most successful [34,35].

The energy filtering idea [36,37], although different from the density-of-states argument, preferentially scatters electrons of lower-energy, which can theoretically lead to higher $S^2\sigma$ as well. Experimental evidence for the filter energy filtering effects was provided in nanogranular PbTe [38] and InGaAs/InGaAlAs superlattices [39]. Other progress was made in improving the electron performance through modulation doping [40], carrier-pocket engineering that produces convergence of different bands in both bulk materials through temperature change [41] and in nanostructures through quantum size effects [42], and anti-resonant scattering [43,44].

In addition to the electron properties as described by the transport coefficients, phonons are dominant heat carriers in most semiconductors and reducing the phononic thermal conductivity has been a major direction in improving ZT. In an isotropic crystalline material, phonon thermal conductivity can be expressed using simple kinetic theory as

$$\kappa_p = \frac{1}{3}\int_0^{\omega_{max}} C(\omega) v(\omega) \Lambda(\omega) d\omega = \frac{1}{3}\int_0^{\omega_{max}} C(\omega) v^2(\omega) \tau(\omega) d\omega \qquad (2.17)$$

where $C(\omega) = \hbar\omega D(\omega) df_{BE}/dT$ is the specific heat per unit frequency interval at frequency ω and temperature T, $D(\omega)$ is phonon density of states per unit volume and per unit frequency interval, $f_{BE}$ is the Bose-Einstein distribution, $v(\omega)$ is the phonon group velocity, $\Lambda(\omega) = v(\omega)\tau(\omega)$ is the phonon mean free path (MFP) and $\tau(\omega)$ is the phonon lifetime. We will discuss how $\kappa_p$ can be reduced in subsequent sections.

## 2.2 Thermoelectric Device Basics

Along the device direction, thermoelectric devices often consist of many pairs of p-type and n-type semiconductors pellets connected electrically in series and thermally in parallel (Fig.4). Taking one pellet in Fig.(4b) working in the power generation mode, Eq. (2.8) can be solved for the temperature distribution along the leg when the two ends are subject to constant temperatures $T_H$ and $T_C$. If we assume one-dimensional transport along the pellet axis with temperature-independent properties, which also leads to a Thomson coefficient of zero, the equation can be solved readily for the temperature distribution. Applying Eq. (2.5) and an energy balance at the two boundaries, we find the heat transfer at the two boundaries is

$$Q_{H,p} = S_p I T_H - \frac{I^2 \rho_p L_p}{2A_p} + \frac{\kappa_p A_p (T_H - T_C)}{L_P} \qquad (2.18)$$

$$Q_{C,p} = S_p I T_C + \frac{I^2 \rho_p L_p}{2A_p} + \frac{\kappa_p A_p (T_H - T_C)}{L_P} \qquad (2.19)$$

where $\rho = 1/\sigma$ is the electrical resistivity, A is the cross-sectional area of the leg and L is the length of leg. The first tem represents Peltier heat, the second term is due to Joule heating and the third term is due to heat conduction. The subscript p denotes properties of the p-type pellet.

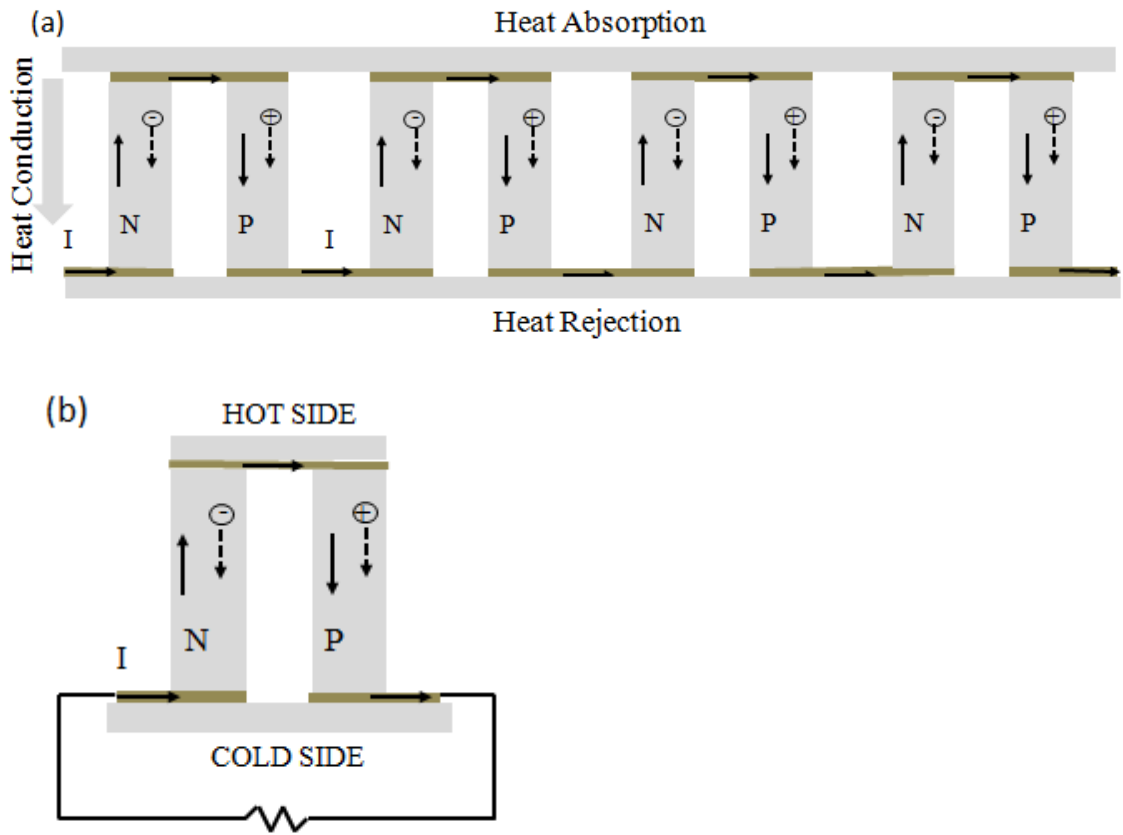

Fig.4 (a) Thermoelectric module with many pairs of p-type and n-type semiconductor pellets connected electrically in series and thermally in parallel. (b) Schematic of thermoelectric power generator shown with one p-n pair.

For a power generation device with both n-type and p-type legs, the electrical resistance of the two legs in series is

$$R = \frac{L_p \rho_p}{A_p} + \frac{L_n \rho_n}{A_n} \tag{2.20}$$

and the thermal conductance of the two legs in parallel is

$$K = \frac{\kappa_p A_p}{L_p} + \frac{\kappa_n A_n}{L_n} \tag{2.21}$$

The power output can be written as

$$W = \left[ \frac{(S_p - S_n)(T_H - T_C)}{R + R_L} \right]^2 R_L \tag{2.22}$$

where $R_L$ is the external load resistance of the output circuit, and the efficiency of power conversion is

$$\eta = \frac{W}{Q_H} = \frac{\left[ \frac{(S_p - S_n)(T_H - T_C)}{R + R_L} \right]^2 R_L}{\frac{(S_p - S_n)^2 (T_H - T_C)}{R + R_L} T_H - \frac{1}{2} \left[ \frac{(S_p - S_n)(T_H - T_C)}{R + R_L} \right]^2 R + K(T_H - T_C)} \tag{2.23}$$

This efficiency is a function of the external load, $R_L$. An optimal is found by setting $\partial \eta / \partial R_L = 0$, which leads to [3]

$$\eta_{max} = \frac{(T_H - T_C)(\sqrt{1 + ZT_M} - 1)}{T_H(\sqrt{1 + ZT_M} + T_C / T_H)} \tag{2.24}$$

when $R_L = R\sqrt{1 + ZT_M}$ with $Z = (S_p - S_n)^2 T_M / (KR)$ and $T_M = (T_H + T_C)/2$.

In the above expression, we see that efficiency is a function of the Carnot efficiency multipled by a factor that depends on $S^2 T / KR$. The larger the latter factor, the higher the efficiency will be. One can maximize this factor by choosing the geometry of the two legs to satisfy the following conditions:

$$\frac{L_n A_p}{L_p A_n} = \left( \frac{\rho_p \kappa_n}{\rho_n \kappa_p} \right)^{1/2} \tag{2.25}$$

And the maximum is

$$(ZT)_{max} = \frac{(S_p - S_n)^2}{(\sqrt{\kappa_p \rho_p} + \sqrt{\kappa_n \rho_n})^2} T \qquad (2.26)$$

The above analysis shows that for a thermoelectric device, one needs to optimize the geometry and use materials with large ZT values to maximize efficiency. True ZT values of a pair are dependent on a combination of the properties of the two legs. For simplicity in the materials search, however, we usually use ZT for individual materials as defined in Eq. (1.1) to gauge their potential for thermoelectric applications.

Similar analysis for thermoelectric cooling leads to the following expression for the coefficient of performance (COP) for refrigeration

$$\phi_{max} = \frac{T_C(\sqrt{1+ZT_M} - T_H/T_C)}{(T_H - T_C)(\sqrt{1+ZT_M} + 1)} \qquad (2.27)$$

and for heat-pumping:

$$\phi_{max} = \frac{T_H(\sqrt{1+ZT_M} - T_C/T_H)}{(T_H - T_C)(\sqrt{1+ZT_M} + 1)} \qquad (2.28)$$

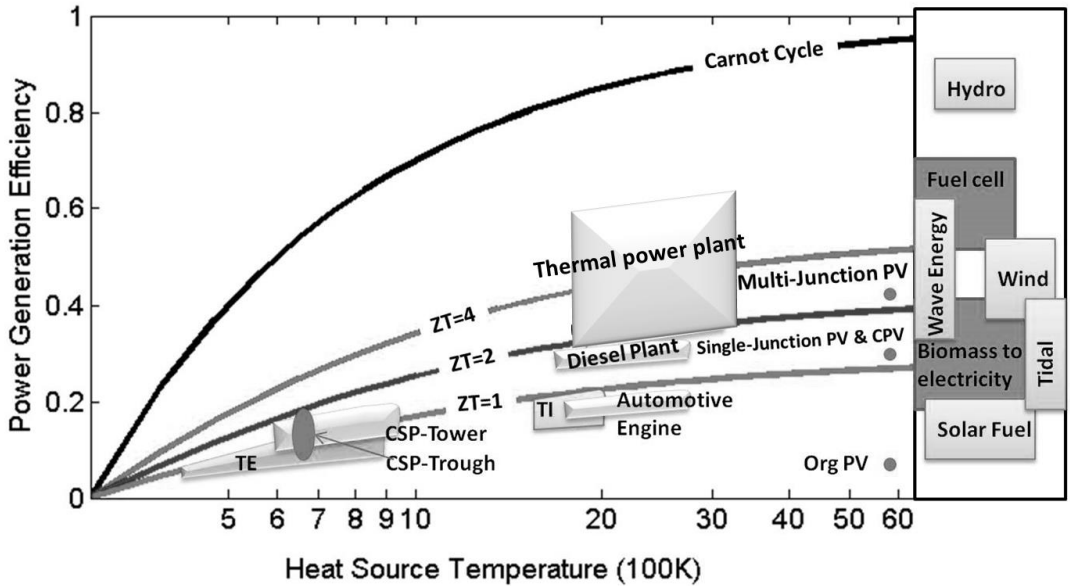

Fig. 5 Power generation efficiency versus temperature of the hot side for different energy conversion technologies where the cold side is assumed to be at room temperature [9]. PV stands for photovoltaic, CSP denotes concentrated solar power, and Org, TE, TI means organic, thermoelectric, and thermionic devices respectively.

Figure 5 plots the efficiency of ideal thermoelectric generators as a function of the hot side temperature at different ZT values, and compares this ideal efficiency with reported efficiency of other technologies. Current materials have ZT values around 1-2, and the efficiency of power generators is not high enough to compete against prime power converters such as steam turbines and internal combustion engines. However, thermoelectric generators have efficiencies comparable to other energy devices such as photovoltaics (certain types only), solar thermal plants, etc. In the cooling area, thermoelectric devices are not competitive against compressor based refrigerators. However, the solid-state nature of the thermoelectric coolers means that they can be made compact. This unique advantage has supported a small market for thermoelectric coolers with applications such as wavelength stabilization for semiconductor lasers, small scale refrigerators for dormitories and on-board refrigeration systems, temperature control of scientific instruments, etc. [45].

It is clear that thermoelectric energy conversion always desires materials with higher ZT values. Large amounts of effort have been devoted to ZT enhancement by improving existing materials and discovering new ones. We would like to emphasize, however, that even with current materials, there is space for innovation and development of thermoelectric technology to take advantage of their solid-state nature, scalability, silent operation, and environmental friendliness. Many times, it is poor heat transfer design that prevents achieving the ideal efficiency as shown in Fig. 5. The heat transfer community can play an active role in advancing thermoelectric technology, by contributing to the understanding of heat transfer physics in materials, and device and system innovation. Our review will emphasize heat transfer physics in materials and device level heat transfer issues.

## 3. Progress in Thermoelectric Materials

Classical thermoelectric materials are based on $Bi_2Te_3$ and its alloys with $Bi_2Se_3$ and $Sb_2Te_3$, PbTe and its alloys with PbSe and SnTe, and SiGe alloys. These materials were identified in the 1950s [46-48]. $Bi_2Te_3$ and PbTe contain heavy elements in the periodic table, which lead to small phonon group velocity and low thermal conductivity. Such heavy elements usually have small bandgaps and large mobility [20]. SiGe alloys, however, have large bandgaps and relatively light masses compared to $Bi_2Te_3$ and PbTe. Ioffe [49] suggested that thermal conductivity can be reduced by alloying, and this strategy has proved to be effective in all thermoelectric materials. These three materials cover a quite wide range of operation temperatures. Currently, Peltier cooling devices on the market are almost exclusively based on $Bi_2Te_3$. Power generation used in NASA space missions relies mostly on SiGe alloys because they have a large bandgap such that the cold side can be operated at high temperatures to reduce the size of radiators and have negligible degradation up to 1300 K, although PbTe had also been used before in the 1960s and early 1970s [50].

These classical thermoelectric materials were discovered in the 1950s. Between the 1950s and the 1990s, progress in ZT improvements was slow. However, in the 1990s, interest in thermoelectric energy conversion was renewed, first in potential refrigeration applications and subsequently more in power generation. New scientific ideas [30,31,36,51], together with investments from the US government, and other governments all over the world led to impressive progress in materials, as we will briefly summarize below.

### 3.1 Bulk Thermoelectric Materials

Alloying remains an active research area. $Bi_2Te_3$ alloys, p-type $Bi_xSb_{2-x}Te_3$ and n-type $Bi_2Te_{1-x}Se_x$ [10], have ZT values around 1 and have been in commercial thermoelectric cooling modules near room temperature and power generation modules for temperatures up to 500 K since their discovery in the 1950s [46]. Group-IV tellurides are typical materials for 600-800 K range. PbTe alloys, such as p-type $PbTe_{1-x}Se_x$, and n-type $Pb_{1-x}Sn_xTe$, were reported to exhibit ZT~1 [2,5] with over 40% thermal conductivity reduction from that of PbTe at 300K. Most recent efforts on p-type $PbTe_{1-x}Se_x$ were able to achieve ZT~1.6-1.8 [20,41,52], along with band engineering to optimize the power factor, although lately a record high ZT of 2.2 is reported by concurrently increasing the electron power factor and reducing the phonon thermal conductivity [53]. Another notable material $(AgSbTe_2)_{1-x}(GeTe)_x$, known as TAGS, has shown ZT>1 for both p-type and n-type, although only p-type with ZT~1.2 has been successfully used in long-term power generators [4,20]. SiGe alloys are candidates for high temperature applications because their peak ZT values locate at > 900 K and they exhibit long reliability. $Si_{0.8}Ge_{0.2}$ possesses ZT~1 for n-type and ZT~0.6 for p-type [54] but recent efforts through nanostructuring has led to peak ZT of 1.3 for n-type [55] and 0.9-1 for p-type [56].

Ideal thermoelectric materials are sometimes summarized as phonon-glass electron-crystals (PGEC). The terminology of PGEC was first put forward by Slack [57] to describe that ideal thermoelectric materials should achieve glass-like lattice thermal conductivity but crystal-like electron transport [6]. One new idea Slack proposed is to reduce phonon thermal conductivity by the so called "phonon rattlers". By introducing guest atoms or molecules into the voids of an open cage structure, the atomic rattlers form phonon scattering centers and interact with a broad spectrum of phonons. The lattice thermal conductivity can thus be substantially reduced. Two representative classes of materials that have received attention are skutterudites and clathrates.

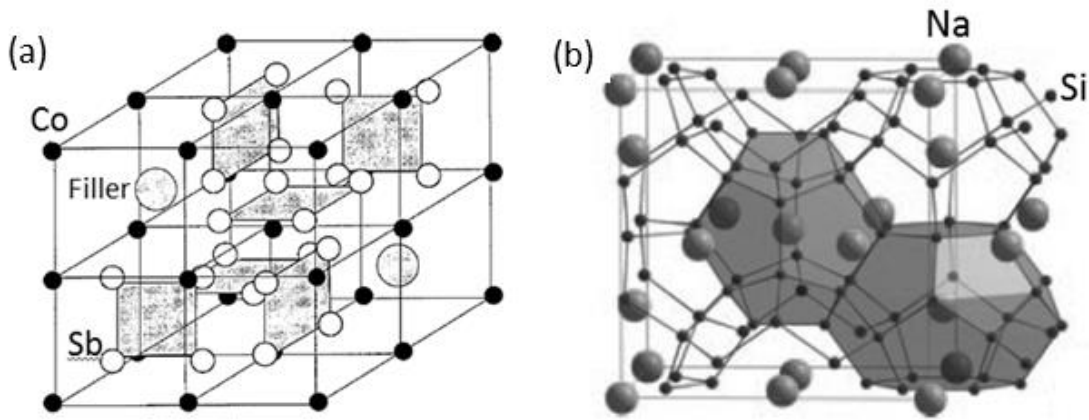

Fig.6 (a) Skutterudites $CoSb_3$ structure where the void cages are filled with filler [5]. (b) Type I clathrates $Na_8Si_{46}$ where Na is the filler [4].

Skutterudites are actively pursued by many groups for potential power generation applications in the 500-900 K temperature range. The crystal formula of Skuetterudites can be written in $MX_3$ where M is Co, Rh or Ir and X is P, As or Sb. The key feature is large empty space. For example, as shown in Fig. 6(a), each unit cell of $CoSb_3$ contains eight pseudo cubes formed by Co and six squares formed by Sb. The voids can be filled by different elements such as rare-earth, alkaline-earth or other heavy atoms [58-61]. Although at the beginning, Slack tried to insert inert atoms such as argon, remarkable reduction in thermal conductivity was demonstrated by introducing rare earth atoms into skutterudites [60] even at a small (5%) fractions [61]. Lattice thermal conductivity of rare-earth filled skutterudites including $Ir_4LaGe_3Sb_9$, $Ir_4NdGe_3Sb_9$ and $Ir_4SmGe_3Sb_9$ was reported to be over an order of magnitude lower than the unfilled $IrSb_3$ [59]. In addition to the rattling effect, the phonon-electron scattering in the case of $Nd^{3+}$ and $Sm^{3+}$ helped reduce lattice thermal conductivity, but the electronic properties also degraded. ZT was thus not improved. Neutral atoms or molecules were suggested as fillers. Recently, the double-filled n-type $Ba_xYb_yCo_4Sb_{12}$ was observed to have an order of magnitude reduction in the lattice thermal conductivity (~0.9 W/mK at room temperature) compared to $CoSb_3$ and was shown to have a ZT~1.4 [62] while triple-filled n-type $Ba_xLa_yYb_zCo_4Sb_{12}$ possessed even lower lattice thermal conductivity (~0.2 W/mK) with a ZT~1.7 [63]. Progress with p-type skutterudites has been slower since filling tends to drive skutterudites strong n-type and the best p-type skutterudites have ZT~1 [64].

Clathrates are another class of compound which has open structures to host loosely bounded guest atoms. Clathrates typically have a large number of atoms in the unit cell and generally have low thermal conductivity (~1 W/mK at room temperature) that is comparable to that of amorphous germanium[2]. Two types of clathrates exist, with Type I more common, as shown in Fig. 6(b). Type I clathrates can be represented by $X_2Y_6E_{46}$ where X and Y are guest atoms encapsulated in two different cages and E is Si, Ge or Sn. A highest ZT~1.4 has been reported for n-type single crystal $Ba_8Ga_{16}Ge_{30}$ [65], while fairly low ZT~0.6 was observed for $Ba_8Ga_{16}Al_3Ge_{27}$ [66].

Another idea to approach PGEC is to look for more complex structures, either large unit cells or complex compositions. Complex crystal structures have more optical phonon branches with low group velocities and consequently low thermal conductivities. Complex structures are known to have exceptionally low lattice thermal conductivities intrinsically [10], such as β-$Zn_4Sb_3$ (0.65 W/mK [67]), the Zintl phase $Yb_{14}MnSb_{11}$ (0.4 W/mK [68]), and thallium chalcogenides ($Tl_9BiTe_6$ ~0.39 W/mK [69], $Ag_9TlTe_5$ ~0.23 W/mK [70]). Rare-earth chalcogenides [20] with $Th_3P_4$ structure have a large number of random vacancies and relatively low thermal conductivity ($La_{3-x}Te_4$ ~0.5 W/mK [10]).

Half-heusler compounds possess a cubic structure consisting of three interpenetrating FCC sublattices and one vacant sublattice, with high thermal stability and environmental friendliness. They exhibit promising power factors when proper doped. Uher *et al.* [71] collected extensive transport property data of pure and doped half-heusler alloys. They unveiled the unusual electron transport properties related to a semimetal-semiconductor transition at low temperatures and the presence of heavy electrons at high temperatures. Remarkably high power factor was observed in n-type alloys. However, half-heusler compounds typically have a relatively high lattice thermal conductivity of ~10 W/mK. By alloying and nanostructuring, the lattice thermal conductivity can be reduced to ~3 W/mK and lead to ZT~1 in p-type $Zr_{1-x}Ti_xCoSb_{0.8}Sn_{0.2}$ [72] and n-type $Hf_{1-x}Zr_xNiSb_{0.99}Sn_{0.01}$ [73]. Later it was also demonstrated that a larger mass difference between Hf and Ti than between Zr and Ti led to more severe thermal conductivity reduction in $Hf_{1-x}Ti_xCoSb_{0.8}Sn_{0.2}$ [74] (~2W/mK) than $Zr_{1-x}Ti_xCoSb_{0.8}Sn_{0.2}$ [72].

Oxides, a relative newcomer to thermoelectric fields, are potentially stable and chemically inert for high-temperature applications. Oxides were believed to be poor thermoelectric materials because of the low carrier mobility arising from the weak orbital overlap and localized electrons until the discoveries of good p-type thermoelectric properties in layered cobaltites $NaCo_2O_4$ [75], $Ca_4Co_3O_9$ [76] and $Bi_2Sr_2Co_2O_9$ [77] with large Seebeck coefficients, low thermal conductivities (<1 W/mK) and ZT~1 at 700-1000 K. This is strikingly against common sense that low mobility materials cannot be a thermoelectric materials. The most promising candidates for *n*-type oxide thermoelectric materials include perovskite-type $SrTiO_3$ [78] and $CaMnO_3$ [79]. Yet their ZT values are still low, as they have rather high thermal conductivities ($SrTiO_3$ ~8 W/mK). By partially substituting Dy for Sr, the thermal conductivity can drop to 3.4 W/mK [80].

### 3.2 Nanostructured Thermoelectric Materials

Besides the efforts to pursue PGEC in bulk materials, one major direction to improve ZT moves towards nanostructuring. Hicks and Dresselhaus [30,31] pioneered the concept that low-dimensional materials can have preferentially modified material properties to enhance ZT due to quantum size effects on electrons and classical size effects on phonons. The power factor benefits from increased electron density of states near the Fermi level due to quantum confinement. In addition, strong phonon scattering (less emphasized in their paper) at interfaces and boundaries was introduced and lead to lower thermal conductivity by simply setting the phonon MFPs to be the film thickness. Venkatasubramanian [51] suggested that low-thermal conductivity in the cross-plane direction of superlattices, as experimentally observed [81], can also lead to higher ZT. These studies opened a new door for thermoelectric materials research. Early work focused more on the unique electronic properties stemming from the low dimensionality. Gradually, it was recognized that a major advantage of nanostructuring is phonon thermal conductivity reduction [82]. This recognition has gone several directions: one is continued studies on individual nanostructures such as nanowires and superlattices, and another is bulk nanostructures.

Initial experimental focus on low-dimensional structures was on superlattices and quantum wells, as suggested by Hick and Dresselhasus [30] and Venkatasubrmanian [51]. Venkatasubramanian reported that p-type $Bi_2Te_3/Sb_2Te_3$ superlattices [51] exhibit an exceptionally high ZT of ~2.4 at room temperature with a remarkably low thermal conductivity of ~ 0.22 W/mK that is lower than $Bi_{0.5}Sb_{1.5}Te_3$ (~0.5 W/mK). Harman[83] reported quantum-dot superlattices made of PbSeTe/PbTe with a ZT~2. We should caution that these data have not been reproduced by other groups. However, it is clear that superlattices have remarkably low thermal conductivity. Many groups reported superlattices of Si/Ge [84-86], GaAs/AlAs [87,88] and PbTe/PbSe [89] have even lower lattice thermal conductivity than their alloy counterparts.

Parallel to 2D structures, Dresselhaus and co-authors also initiated work on nanowire structures, first demonstrating semimetal to semiconductor transitions as predicted by theory [90]. Later, remarkable lattice thermal conductivity reduction in different nanowires was reported. For example, GaN [91,92], $Bi_2Te_3$ [93], InSb [94], InAs [95] and Bi [96] nanowires have notable thermal conductivity reduction from phonon-boundary scattering. The most significant claims, however, arise from silicon nanowires. Bulk Si with a high thermal conductivity ~150 W/mK is a poor thermoelectric material with ZT ~ 0.01 at room temperature. By dramatically suppressing the lattice thermal conductivity without much disturbance to the power factor, Si nanowires have shown promising ZT values. Hochbaum et al. [97] reported ZT = 0.6 at room temperature for rough Si naowires and Boukai et al. [98] claimed ZT ~ 1 at 200K for Si nanowires, where the thermal conductivities were both reduced to below 2 W/mK.

Other than nanowires made of one pure material, core-shell nanowires [99] or alloyed nanowires [100] may further reduce thermal conductivity. Nanotubes are another 1D nanostructure receiving incremental attention for thermoelectric applications. In fact, nanotubes may reach even lower lattice thermal conductivities than nanowires due to the additional phonon scattering from the holy structure. For example, $Bi_2Te_3$ nanotubes [101] showed notably low lattice thermal conductivity (~ 0.3 W/mK) and ZT ~1.

Guided by the understanding that the reduced thermal conductivity in superlattices was due to diffuse interface scattering rather than phonon band structure modification, Yang and Chen [102] suggested that random 3D nanostructures can be exploited to achieve high ZT. At the same time, experiments were starting to emerge with improved ZT values in random 3D nanostructures [103]. 3D nanocomposites refer to the format of nanoinclusions such as nanoparticles or nanowires embedded in a host matrix or a heterostructure geometry with nanostructures adjacent to each other. Nanocomposites that do not require any specific morphology are practical for mass fabrication and easy to incorporate into commercial devices. In fact, nanograined materials from the same type usually show desirable thermal conductivity reduction, although these materials often possess additional internal structures such as nanoprecipitates [104]. To preserve the electronic properties, one should also bear in mind that the bandgap between the nanophase and matrix or different nanophases cannot be large and the electrostatic potential due to impurity or oxidation should be small.

Two major approaches to make bulk nanocomposites are hot-pressing the ball-milled nanopowers or self-forming inhomogeneities by phase segregation. Despite different approaches, both can reduce the lattice thermal conductivity due to increasing phonon-interface scattering. Ball-milling introduces nanoscale grains within the samples. For example, Ren group [8] have grown a variety of thermoelectric nanocomposites with promising ZT values, including the striking ZT of 1.4 for nano-$Bi_xSb_{2-x}Te_3$ [105,106]. PbTe-based systems have been shown to spontaneously form nanoinclusions within a PbTe rocksalt matrix during thermal processing. For example, in a PbTe-PbS system shape-controlled cubic PbS nanostructures were observed in the PbTe matrix. The considerably low lattice thermal conductivity (~0.7 W/mK at 800 K) was ascribed for a high ZT (~1.8 at 800 K) of p-type PbTe-PbS 12% [107]. $NaPb_mSbTe_{2+m}$ (SALT-*m*) is also a high-performance p-type material (ZT~1.7 at 650 K). The impressive ZT is also attributed to the very low lattice thermal conductivity (0.74 W/mK at 300 K and 0.55 W/mK at 650 K) caused by pervasive nanostructuring [108].

Nanostructured alloys are proven to be powerful in lowering the thermal conductivity by taking advantage of phonon scattering at different length scales. For example, ErAs nanoparticles in a $In_{0.53}Ga_{0.47}As$ matrix [109], nanocrystalline $Bi_xSb_{2-x}Te_3$ [105,106] and $Si_{0.8}Ge_{0.2}$ [55], and nanostructured $Bi_{2-x}Cu_xS_3$ [110] have been demonstrated to possess the thermal conductivity 30%-50% lower than the alloy limit. The same idea can also be extended to nanostructured phonon rattling materials. For example, nanostructured $Yb_{0.3}Co_4Sb_{12+y}$ was reported to have a ZT~1.3 at 800K [111]. InSb nanophase with grain sizes ranging from 10-80nm in $In_xCe_yCo_4Sb_{12}$ lead to a notable depression of lattice thermal conductivity and a ZT~1.43 at 800 K [112].

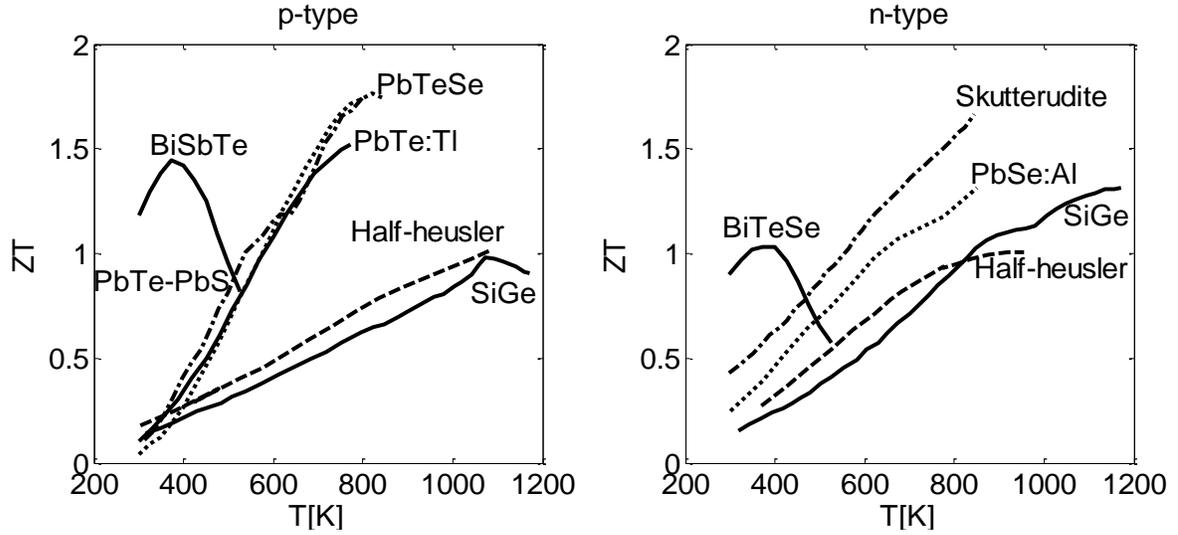

Fig. 7 Figure of merit, ZT, of selected state of the art thermoelectric materials versus temperature for (a) p-type and (b) n-type. (p-type: $Si_{0.8}Ge_{0.2}$[56], $Bi_{0.5}Sb_{1.5}Te_3$[105], 2% Tl-doped PbTe [34], Na-doped $PbTe_{0.85}Se_{0.15}$ [41], Na-doped PbTe-PbS 12% [107], half-heusler $Hf_{0.8}Ti_{0.2}CoSb_{0.8}Sn_{0.2}$ [74]; n-type: $Si_{0.8}Ge_{0.2}$ [55], multi-filled skutterudies $Ba_xLa_yYb_zCo_4Sb_{12}$ [63], half-heusler $Hf_{0.75}Ti_{0.25}NiSn_{0.99}Sb_{0.01}$ [73], 1% Al-doped PbSe [35], $Bi_2Te_{2.7}Se_{0.3}$ [113])

Figure 7 summarizes ZT values for different materials (plots both n-type and p-type). Considering that the highest ZT values before the 1990s were around 1, progress in thermoelectric materials is indeed very impressive covering a wide temperature range. Many improvements come from reduced lattice thermal conductivity, which will be discussed in the next few sections. In the following sections, we will first discuss heat conduction mechanisms in bulk crystals (Sec.4), and then move on to discuss strategies and corresponding mechanisms used in reducing thermal conductivity in both bulk materials (Sec. 5) and nanostructures (Sec. 6).

## 4. Heat Conduction in Bulk Thermoelectric Materials

In typical semiconductors, the electronic contribution to thermal conductivity is usually masked by the phononic contribution. However, the electronic contribution to thermal conductivity becomes appreciable in materials with high ZT values. In this section, we will discuss heat conduction mechanisms in bulk crystals, including the contributions of both electrons and phonons.

### 4.1 Heat Conduction by Phonons

Understanding heat conduction in crystalline solids started only when the quantum theory of lattice vibration was developed and the phonon concept was established. The Einstein model [114], and subsequently Debye model [115] based on Planck's quantization, were able to account for the temperature dependence of the lattice specific heat observed experimentally (although Einstein model did not provide the exact temperature trend at low temperatures), leading to the harmonic oscillator model, and the corresponding Bose-Einstein distribution for the basic quanta of lattice vibration---a phonon. In an ideal case, the harmonic interatomic potential gives an infinite thermal conductivity. In a real crystal, however, the anharmonic part in interatomic potential results in scattering among phonons and a finite thermal conductivity. The phonon scattering originating from the anharmonic interatomic potential can be classified as either a normal process which conserves crystal momentum or an Umklapp process which does not. It was recognized that the Umklapp process creates a resistance to heat flow while the normal process only redistributes phonons. Peierls [116] extended the Boltzmann formulation for phonon transport taking into consideration only the Umklapp process for phonon scattering. Under the relaxation time approximation and assuming isotropic group velocity and phonon lifetime, the Boltzmann equation led to the kinetic theory expression for thermal conductivity as given in Eq. (2.17). Callaway [117] further modified the theory by accounting for both the normal and Umklapp scattering processes, assuming that normal processes lead to a displaced Bose-Einstein distribution and indirectly affect the Umklapp processes. The derived expression is similar to Eq. (2.17) but with a modification to phonon lifetimes. The quantities in Eq. (2.17) that determine the lattice thermal conductivity, however, are frequency dependent and difficult to obtain. Different approximations were made to perform the thermal conductivity integral.

The Debye approximation that assumes a linear relation between the phonon frequency and the propagation wavevector is often used for the phonon group velocity and density of states, based on the fact that the Debye approximation has been very successfully used to explain the specific heat of crystalline materials. This left the phonon lifetime unknown. In the 1950s, Klemens [118] derived expressions using the quantum perturbation theory, *i.e.*, Fermi's golden rule, for phonon lifetimes due to different scattering mechanisms. The expressions obtained are again based on the Debye approximation and contains unknown parameters. The experimental values of thermal conductivity, speed of sound, and specific heat are often used to fit the unknown parameters in the Debye model and in the lifetimes. Such fittings usually work well at low temperatures. Deviations at high temperatures stimulated more refined models of dispersion, such as the one developed by Holland [119] that used two different linear dispersions to better represent the rapid flattening of transverse acoustic phonons in FCC crystals such as GaAs and Si. A typical fitting curve is shown in Fig. 8.

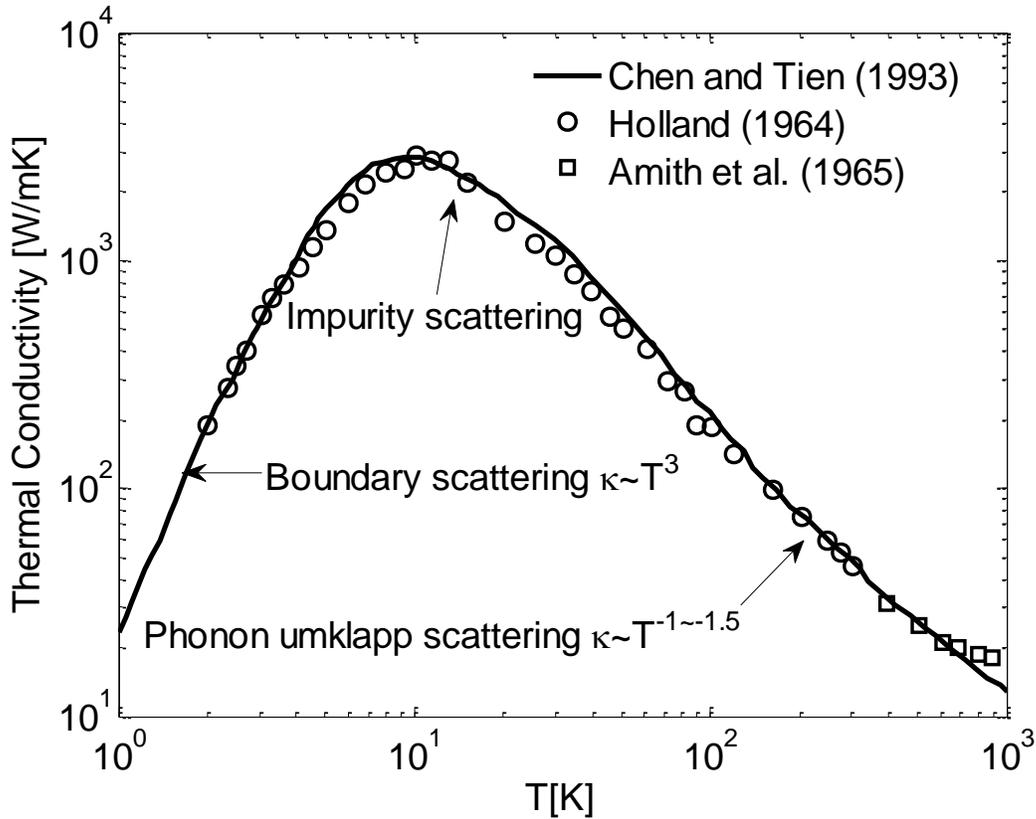

Fig. 8 Calculated thermal conductivity of GaAs, based on a model that considers phonon dispersion and contributions from different phonon branches (solid line) compared with two sets of experimental data (symbols). Replotted based on Ref. [120].

The shape of the temperature dependence of thermal conductivity as in Fig.8 is universal among crystalline solids, and different trends across the temperature range represent different phonon scattering mechanisms. At high temperatures, thermal conductivity usually decreases with increasing temperature as $T^{-n}$, with theoretically n=1 although practically n=1-1.5. This is because at high temperatures, phonon specific heat is a constant according to the Pettit-Delong law, and phonon energy increases linearly with temperature, i.e. the number of phonons increases linearly with temperature. As the scattering rate is proportional to the number of phonons, the thermal conductivity decreases with increasing temperature. At low temperatures, the thermal conductivity is usually proportional to $T^3$, because in this regime, phonon-phonon scattering is weak and phonon MFPs are longer than the size of the sample. Phonons scatter more frequently with the boundaries and the phonon MFP is effectively equal to the sample size, and is independent of frequency. In this case, thermal conductivity is proportional to the specific heat and hence with the $T^3$ behavior. This size effect was first discovered by De Haas and Biermasz [121], and explained by Casimir [122] (sometimes called the Casimir regime). Casimir assumed that surfaces of samples are rough, and scatter phonons diffusely, although latter studies also investigated partially diffuse and partially specular surfaces [123]. In between low and high temperatures, impurity scattering is usually important, and the peak value of thermal conductivity depends sensitively on impurity concentrations [124].

For thermoelectric applications, the question of what is the minimum achievable thermal conductivity was asked. Slack [125] proposed that the minimum could be achieved when phonon MFPs are equal to the full-wavelength of phonons. Based on this assumption, the minimum thermal conductivity can be calculated. Cahill and co-workers [126] argued that instead the minimum MFP should be a half-wavelength, which led to a half of the minimum predicted by Slack.

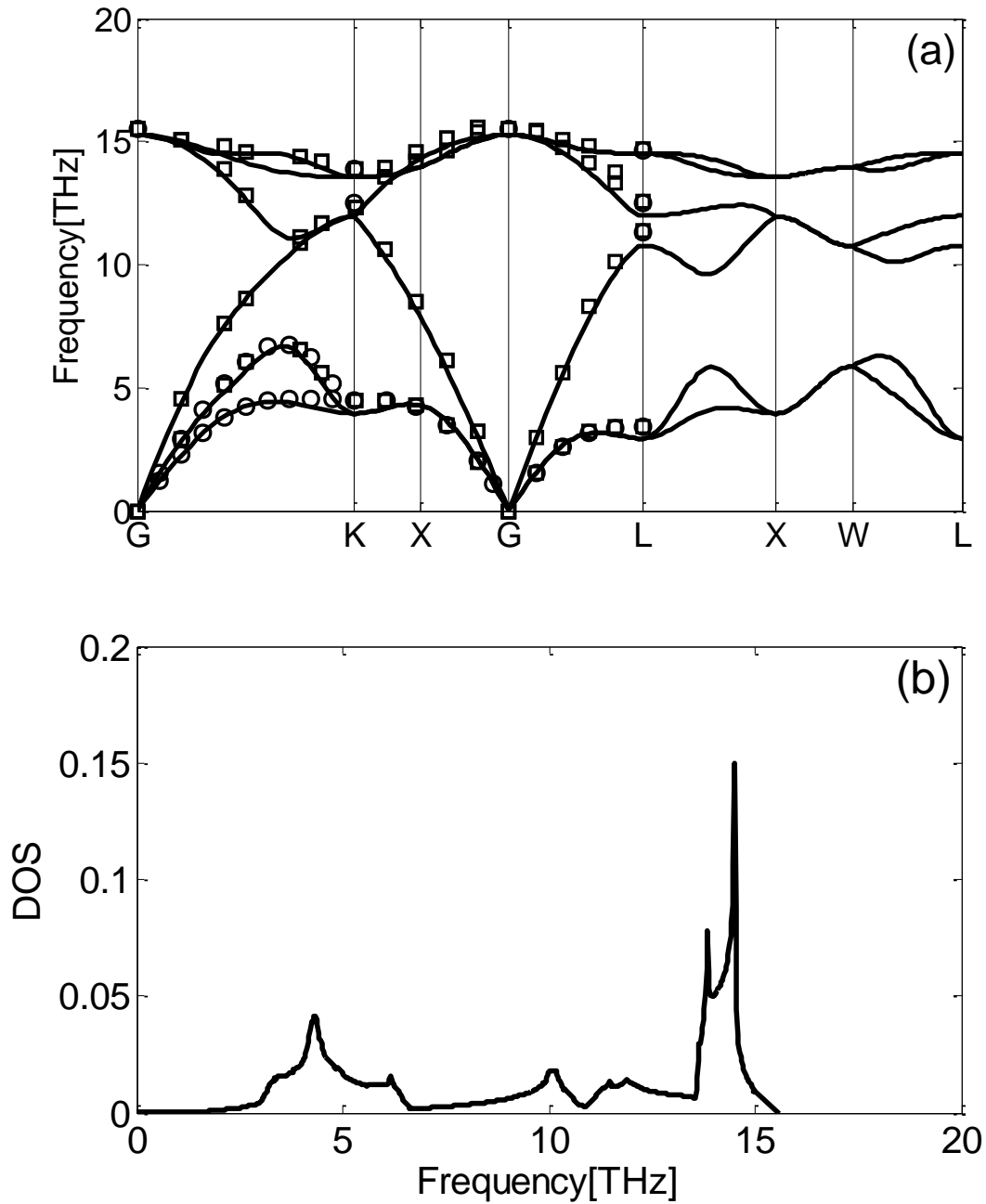

Fig. 9 (a) Phonon dispersion (b) Phonon density of states of Si from first-principles calculations (solid lines) compared with two sets of experimental data (symbols). Replotted based on Ref.[127].

Although the approaches pioneered by Klemens were successful in explaining the trend of experimental observations, quantitative details of phonon transport may be far from complete, especially in terms of the phonon lifetimes and MFPs. The phonon dispersion in a real solid, as shown in Fig. 9(a) for silicon, differs significantly from the Debye model in several aspects. First, optical phonons exist in the crystal, and these phonons have much lower but non-zero group velocities. Second, although low frequency acoustic phonons follow a nearly linear dispersion, high frequency acoustic phonons near the boundaries of the first Brillouin zone are highly dispersive, and their group velocities are usually significantly lower than those represented by the Debye model. Third, due to these differences, the density of states of real crystals usually deviates significantly from the $\omega^2$ law as in a Debye crystal, as shown in Fig.9 (b). The reason that the Debye model together with Klemens' treatment on scattering can fit experimental data is because the thermal conductivity integral is quite forgiving. In fact, one can use different sets of parameters to fit the experimental data based on different approximations. Extracting the exact phonon transport properties, especially phonon lifetimes and MFPs, was thus unsolved.

Since 1980s, efforts started on calculating thermal conductivity of crystals using molecular dynamics (MD) simulations [128-134]. In classical MD, the approximate trajectories of each individual atom within the simulation domain are tracked based on an empirical interatomic potential and Newton's second law. There are two prevailing methods used to study heat transport: one is equilibrium molecular dynamics (EMD) and the other is nonequilibrium molecular dynamics (NEMD). EMD is suitable for transport properties, whereas NEMD simulates actual transport processes. EMD first obtains the history of individual particles in an equilibrium system, from which the transport properties are extracted on the basis of linear response theory. For example, the Fourier transform of the velocity autocorrelation yields the phonon density of states [135]. The thermal conductivity is calculated from the autocorrelation of the instantaneous heat flux through the Green-Kubo formula [136-139]. For NEMD, one can either impose a temperature difference to calculate the heat flux [140,141], or impose a heat flux to calculate the resulting temperature distribution [142,143], which simulates the actual transport processes. The thermal conductivity is then determined by the Fourier's law. Although both EMD and NEMD can give thermal conductivity, they have their own advantages and disadvantages. The NEMD methods are relatively easy to implement and are usually faster than the EMD methods because the latter requires the calculation of the autocorrelation function, which can take a long time to converge. In addition, the EMD method may involve the artificial autocorrelation caused by often-used periodic boundary condition. However, NEMD also suffers from several drawbacks. Firstly, the statistical foundation of NEMD is not as soundly established as that of EMD [26]. Second, the finite simulation size in NEMD might be shorter than the MFP of some phonon modes, leading to artificial size effects with boundaries imposed at the heat reservoirs. Third, a large temperature difference is applied across a small simulation domain.

To extract the phonon lifetimes and MFPs, the phonon spectral energy density analysis [144-148] has been applied in combination with EMD. The key idea is to project the atomic displacements onto normal mode coordinates and determine the phonon lifetimes by tracing the temporal amplitude decay of each mode or fitting the width of spectral energy density peaks in the frequency domain. To apply this approach, one needs to perform separate analysis from the traditional thermal conductivity calculations using MD, and utilize eigenvalues and eigenvectors from harmonic lattice dynamics calculations.

As a widely adopted simulation tool, MD simulations do not require any *a priori* knowledge of the phonon transport properties, are straightforward to implement, and automatically include the temperature-dependent anharmonicity. Yet MD is only rigorously applicable to solids above the Debye temperature for being entirely classical that assumes each vibration mode is equally excited. At low temperatures, quantum correction to temperature is needed. Moreover, the empirical potentials used in classical MD can cause the thermal conductivity to deviate significantly from the experimental data [149].

Conversely, first-principles calculations where the interatomic potential is obtained directly from the electron charge density via density functional theory (DFT) without any adjustable parameters provide a most reliable way of computing the lattice thermal conductivity. *Ab inito* MD simulations [150,151] use the forces computed on-the-fly by DFT and are computationally expensive, though. The alternative way is to extract the interatomic force constants from DFT calculations of limited atomic configurations and conduct further calculations. After DFT calculations, one can either obtain the force constants from reciprocal space based on the density functional perturbation theory (DFPT) [152,153] or from real space by fitting the force-displacement data in a supercell with a polynomial potential [127,154]. Despite that the real space approach is simpler yet less precise, both approaches are accurate enough to reproduce the experimental results of lattice thermal conductivity as shown in Fig. 10(a). Broido *et al.* [152] calculated the intrinsic lattice thermal conductivity of Si and Ge using reciprocal space approach and obtained excellent agreements with experimental data. The reciprocal approach was later applied to $Si_xGe_{1-x}$ alloys and superlattices by Garg *et al.* [155,156], and PbSe, PbTe and $PbTe_{1-x}Se_x$ by Tian *et al.* [157]. Esfarjani *et al.* started the real space approach with Si [127], and then extended to half-Heusler [158], PbTe [159] and GaAs [160]. Once the harmonic and anharmonic force constants are obtained, one can either perform MD simulations based on developed Taylor expansion potential [158] or employ lattice dynamics calculations: first obtain the vibrational eigen modes based on the harmonic part of the potential, then compute the scattering rates of each mode by treating the anharmonic potential as a perturbation using Fermi's golden rule, and solving the Boltzmann transport equation (BTE) to find the thermal conductivity [127,152,155-160]. Although MD simulations based on fitted potentials have more flexibility to directly obtain the lattice thermal conductivity on complicated structures, lattice dynamics calculations can produce the detailed phonon transport properties without extra efforts because the thermal conductivities are the integrated quantities over the first Brillion zone using either the solution of BTE under single-mode relaxation time approximation [127,155,157-161] or based on an iterative solution of the integral BTE [152].

Understanding fundamental phonon transport properties is crucial for further reducing the thermal conductivity and enhancing ZT. The thermal conductivity accumulation as a function of phonon MFPs from first-principles calculations (BTE+Lattice dynamics) is shown in Fig. 10(b). The phonon MFPs in a material span a wide range of magnitudes, from nanometers to microns. Different materials possess different MFP distributions that can provide insight on the required length-scale of nanostructuring to effectively lower the thermal conductivity. For example, while nanostructures of 100nm can efficiently reduce the thermal conductivity of Si and GaAs, nanostructures of less than 10nm are needed for PbTe and PbSe. While the accumulation curve provides intuitive estimation on nanostructuring sizes, one should bear in mind that the characteristic length, L, of the nanostructures should not directly equate the cutoff MFPs. In other words, not all the phonons with MFPs > L are scattered and even if scattered, they still contribute to the thermal conductivity, although to a lesser extend as limited by the new MFPs.

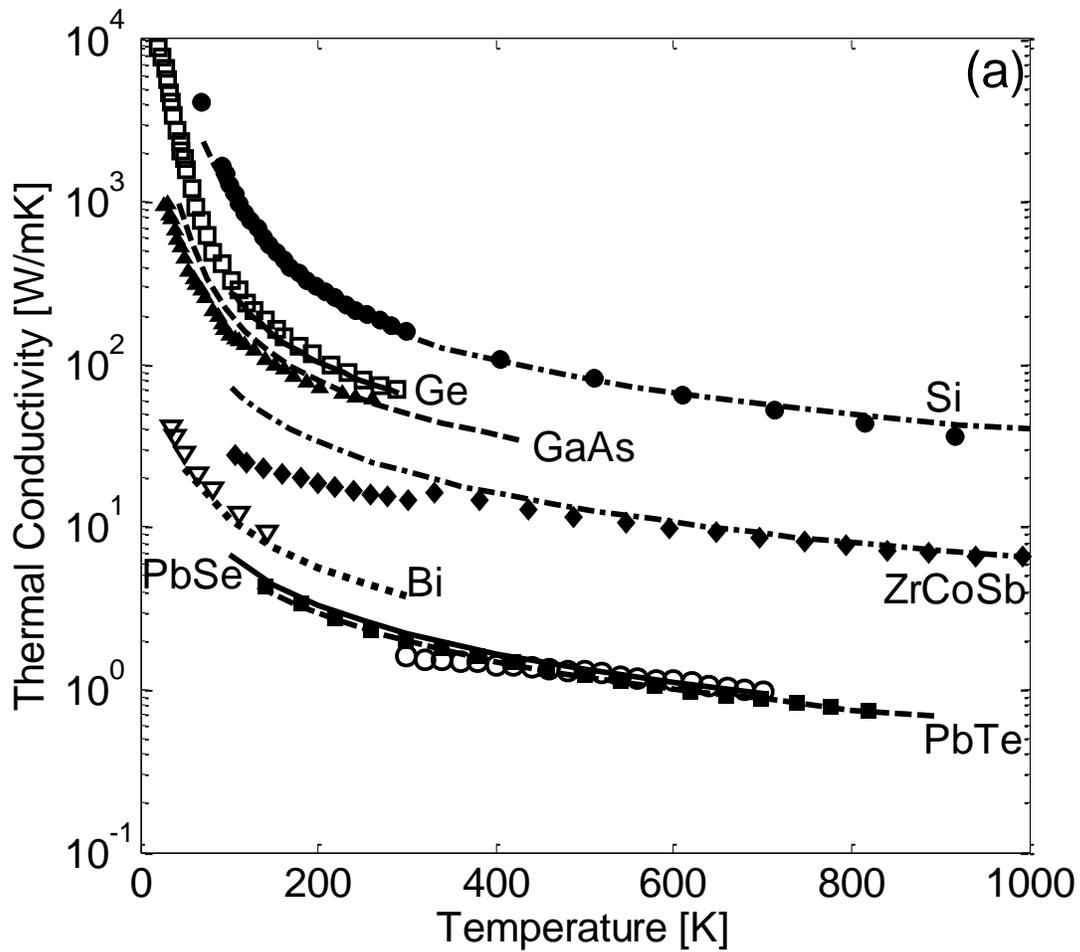

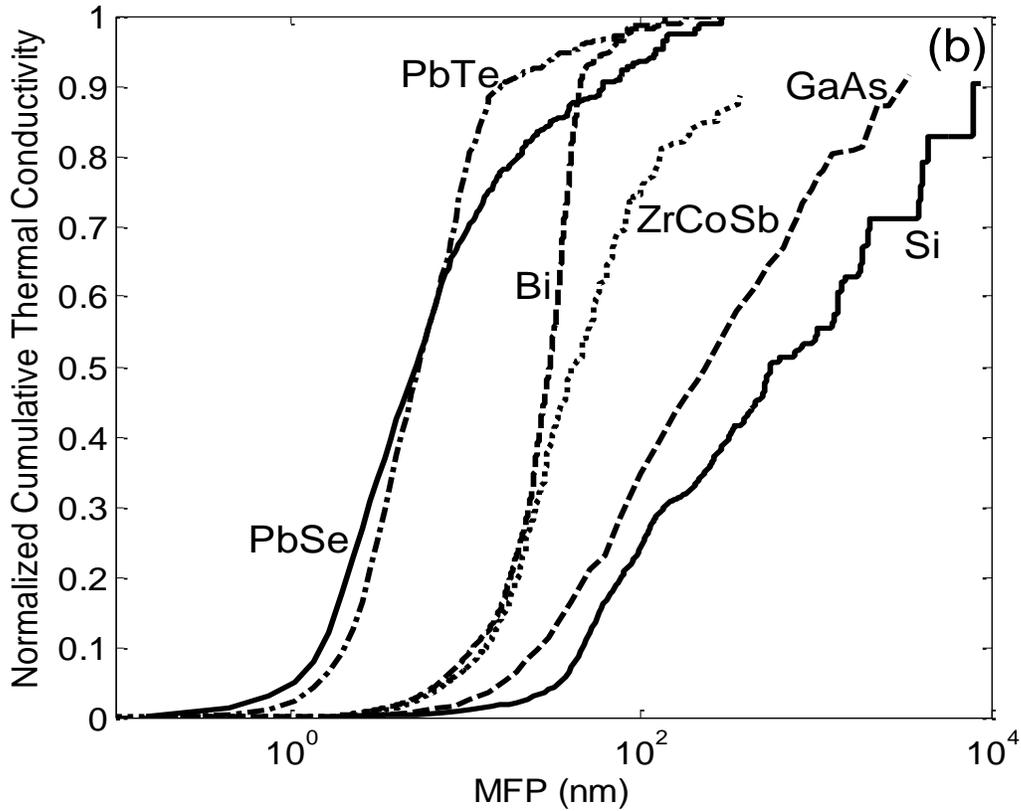

Fig. 10(a) Thermal conductivity versus temperature from first-principles calculations (lines) and experimental measurements (symbols), and (b) Normalized cumulative thermal conductivity as a function of phonon MFPs at 300K from first-principles calculations for different bulk crystalline materials, Si [127], Ge [152], GaAs [160], ZrCoSb [158], PbTe [159], PbSe [157] and Bi [162].

Experimental determination of detailed phonon transport properties is more challenging. Encouraging progress is emerging, though. Using Raman spectroscopy [163,164], inelastic neutron scattering [165] or inelastic X-ray scattering [166], the phonon lifetimes can be extracted from peak width. However, Raman spectroscopy can only probe certain first-order zone center or second-order zone edge modes while the other two measurements only work on bulk single crystals as one needs to distinguish each incident wavevector of neutrons or X-rays. Recent developments in different laboratories have opened new ways to directly determine the thermal conductivity integral in the form of integration over MFPs [144,167]. In time-domain thermoreflectance (TDTR) measurements, Koh and Cahill [168] observed modulation frequency dependence in alloy-based samples. They assumed that phonons with MFPs longer than the thermal penetration traverse ballistically and do not contribute to the measured thermal conductivity which is fitted to a diffusive model. BTE simulation, nevertheless, cannot explain their experimental results [169]. Minnich *et al.* [161] mapped out the accumulative thermal conductivity with respect to phonon MFPs using TDTR by varying the pump beam sizes as shown in Fig.11.

His results match reasonably well with the accumulation curve from first-principles calculations. The key idea is that the heat flux from a heat source will be smaller than Fourier's law's prediction if some phonon MFPs are longer than the heater size such that their transport is ballistic [170]. This was confirmed experimentally using ultrafast coherent soft x-ray beams to probe heat flow from nanolines to a substrate [171]. Transient thermal grating experiments can also excite the quasi-ballistic regime by controlling the grating spacing and the MFP information can be inferred [172]. By numerically inverting the integral function of the total thermal conductivity, Minnich [173] managed to reconstruct the thermal conductivity accumulation curve as a function of MFPs from the measured data. Efforts to push characteristic experimental lengths down to the nanoscale in order to collect the whole range of MFPs and observe the size dependence at elevated temperatures would bring the experimental demonstration to a more complete level [174].

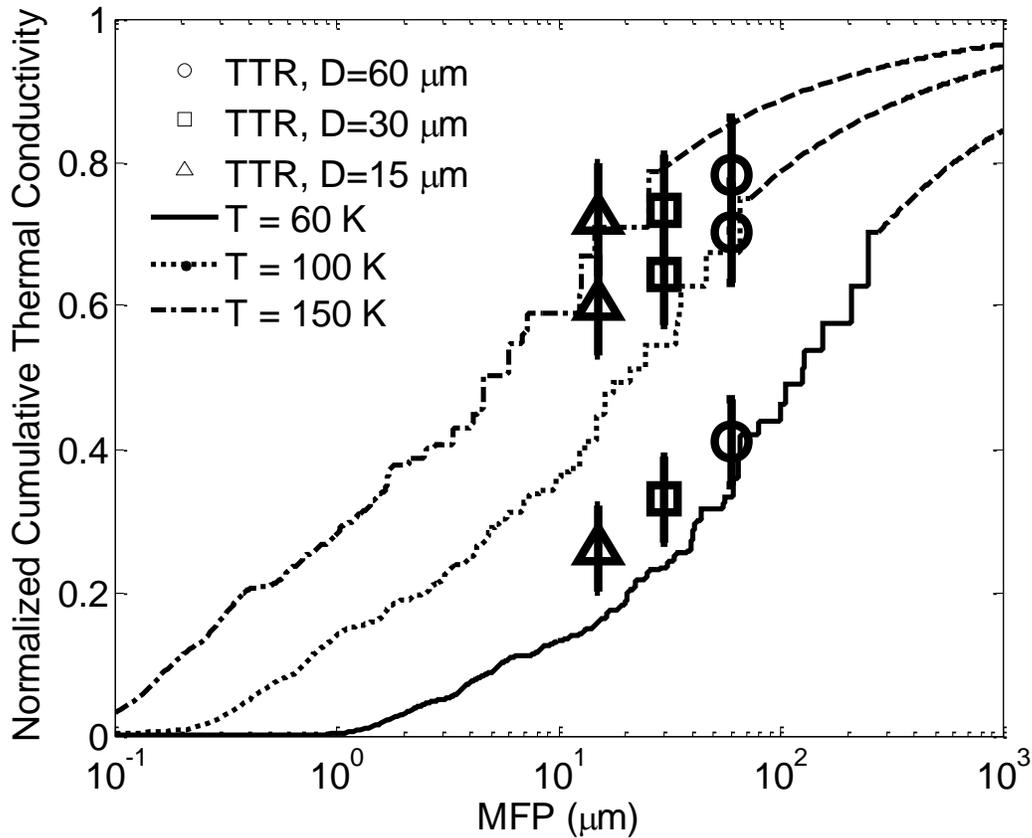

Fig. 11 Thermal conductivity accumulation from experimental measurements (symbols) and first-principles calculations (lines) of natural silicon versus MFPs. Replotted based on Ref. [161]

Detailed phonon transport properties studied both theoretically and experimentally not only provide insights on the effectiveness of nanostructuring to reduce thermal conductivity as discussed above, but also deepen the understanding of transport processes. For example, lead chalcogenides have a high-symmetry rock-salt structure yet surprisingly low thermal conductivity (~2 W/mK) at room temperature. This has intrigued studies on the origin of low thermal conductivity in PbTe and PbSe [157,159,165].

From both first-principle calculations and inelastic neutron scattering experiments, it was found that the strong coupling between acoustic and optical phonon modes and the soft TA modes, in addition to the heavy mass, were the reasons for low thermal conductivity. The phonon transport physics in pure bulk materials thus serves as the fundamental platform to further explore the heat conduction in the more complicated structures.

## 4.2 Heat Conduction by Electrons

In addition to phonons, electrons also carry heat. Most good thermoelectric materials are heavily doped semiconductors to achieve high electrical conductivity. Hence, heat conduction by electrons cannot be neglected in thermoelectric materials. The electronic contribution to thermal conductivity can be calculated from Eq. (2.7). As all the transport coefficients are related to the scattering rate, the electronic thermal conductivity can be related to the electrical conductivity through the Wiedemann-Franz law,

$$\kappa_e = \mathcal{L}\sigma T \tag{4.1}$$

where $\mathcal{L}$ is the Lorenz number. The Wiedemann-Franz law was originally developed for metallic systems and its use and limitation in semiconductors will be discussed later.

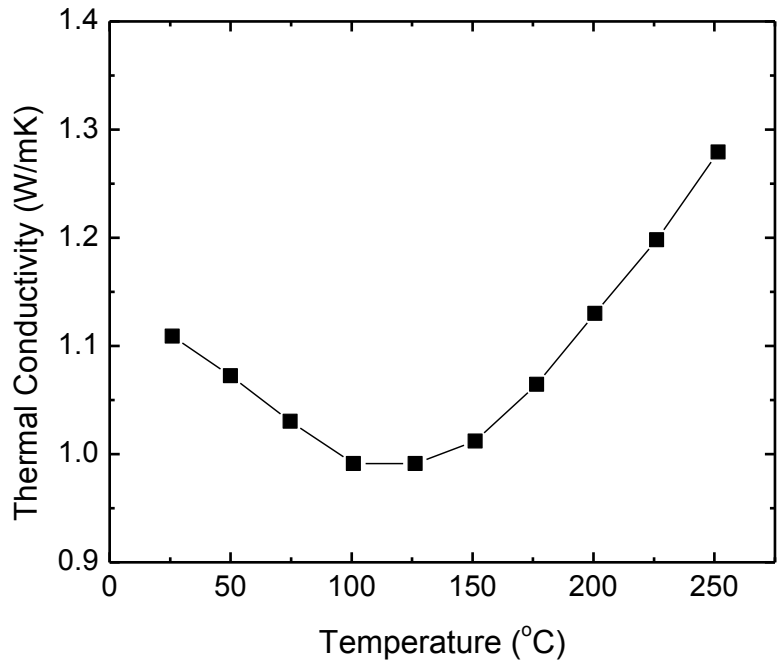

Fig. 12 Thermal conductivity of BiSbTe nanocomposite with temperature [105]

Electronic contribution to heat conduction can be easily observed in the change of total thermal conductivity changes with temperature. If phonons are the predominant heat carrier, the total thermal conductivity should decrease as discussed before. As shown in Fig.12, thermal conductivity in thermoelectric materials typically increases after a certain temperature. This behavior shows that a remarkable portion of heat is carried by charges in that temperature range. As temperature increases, holes and electrons are thermally excited, and carry more heat.

The electronic contribution to thermal conductivity contains two parts. One is due to each individual electrons/holes as determined by the Wiedemann-Franz law. The other is due to energy difference across the bands, which is called the bipolar contribution. The total electronic thermal conductivity can be expressed as [29]

$$\kappa_e = \kappa_n + \kappa_p + \frac{\sigma_n \sigma_p}{\sigma_n + \sigma_p}(S_n - S_p)^2 T \quad (4.2)$$

where $\kappa_e$, $\sigma$, and $S$ represents thermal conductivity, electrical conductivity, and Seebeck coefficient, respectively. The subscript $n$ and $p$ denote electrons and holes. In the expression above, third term shows bipolar heat conduction. In the nondegenerate limit and under constant relaxation time approximation,

$$S_n - S_p = \frac{k_B}{e}\left[\frac{E_f - E_c}{k_B T} - \frac{5}{2}\right] - \frac{k_B}{e}\left[\frac{E_f - E_v}{k_B T} - \frac{5}{2}\right] = -\frac{k_B}{e}\left[\frac{E_g}{k_B T}\right] \quad (4.3)$$

where $E_g$ denotes band gap energy. It is clear that $S_n - S_p$ is proportional to the energy across the bandgap. The bipolar contribution to thermal conductivity is significant at high temperatures when the semiconductor becomes intrinsic such that both $\sigma_n$ and $\sigma_p$ are large. At low temperatures, the prefactor in the third term of the Eq. (4.2) is roughly the ratio of the electrical conductivities of the minority carrier to the majority carrier, which is small, leading to negligible bipolar contributions to the total thermal conductivity.

Reducing electronic and bipolar thermal conductivity would improve ZT, but it remains a challenging problem due to strongly coupled transport characteristics as we discussed in Sec.2. One possible way, having not been demonstrated experimentally, is to modify the density-of-states near the Fermi level to make a delta function shape [175-177]. The maximal ZT condition as discussed in these two theoretical papers, although not stated explicitly, actually corresponds to the limit when the Lorenz number is zero, i.e., the electronic thermal conductivity is zero but the electrical conductivity itself is nonzero. This is possible since Eq. (2.7) shows that the electronic contribution to thermal conductivity contains all three transport coefficients.

To reduce the bipolar effect, one possible way can be devised from the expression of the bipolar thermal conductivity. From the expression, if we can preferentially deteriorate minority carrier's electrical conductivity or Seebeck coefficient, the bipolar contribution can be effectively suppressed. In the past, reduction in the bipolar thermal conductivity was observed in nanocomposite $Bi_xSb_{2-x}Te_3$, which was presumably due to preferential scattering at the grain boundaries [105]. This preferential scattering is advantageous in that it will not affect much majority carrier transport and the power factor. However, further studies are needed to reveal what kind of potential shape can preferentially scatter electrons and how we can control the interfacial potential.

Even though the electronic heat conduction is considerable in thermoelectric materials, it is not easy to quantify the electronic contribution to total heat conduction. The most widely used method to estimate electronic thermal conductivity is to use the Wiedemann-Franz law. As mentioned above, the Wiedemann-Franz law was derived for metals. For metals, the Lorenz number is a constant and can be derived as

$$\mathcal{L} = \frac{\pi^2}{3}\left(\frac{k_\mathrm{B}}{e}\right)^2 = 2.45 \times 10^{-8} \; \mathrm{W\Omega/K^2} \qquad (4.4)$$

However, in semiconductors, the Lorenz number depends on carrier density and electron scattering [178]. More importantly, the Wiedemann-Franz law cannot describe the bipolar thermal diffusion correctly since the band gap is not included in its derivation.

Instead of the Wiedemann-Franz law with the constant Lorenz number for metals, one can measure several electronic transport properties, deduce electronic transport parameters, and then calculate the electronic thermal conductivity [179,180]. One example is to modify the Lorenz number in Wiedemann-Franz law based on measured electron transport properties, such as electrical conductivity and the Hall coefficient to determine the Fermi level, and then compute the electronic contribution to thermal conductivity using methods outlined in Sec. 2. Even though deducing electronic thermal conductivity from other electronic transport properties is more rigorous than using Wiedemann-Franz law with a Lorenz number for metals, still it has many assumptions. One conduction or valence band is often used to reduce the number of fitting parameters. This one band approximation is not applicable near ZT peak temperature, as we discussed before. Furthermore, if multiple valleys exist in a band, which can be observed in $Bi_2Te_{3-x}Se_x$ [181] and $PbTe_{1-x}Se_x$ [182], the one band approximation would not be reasonable. In addition, there exist more uncertainties originating from many other assumptions such as a simple energy dependent scattering model and non-parabolic band structure model. Another way to measure electronic thermal conductivity more directly is to measure thermal conductivity under low temperatures and high magnetic fields. In these cases, the electronic heat conduction is highly suppressed and the measured thermal conductivity mostly comes from phonon contribution [183,184]. However, this method is limited only for low temperatures, and requires a large facility for high magnetic fields. In addition, if a material has large thermomagnetic effects like Bi and $Bi_2Te_3$, then a careful correction for thermomagnetic effects should be also included [183]. Therefore, characterizing electronic heat conduction is not straightforward, and a better method should be developed to understand and control electronic heat conduction.

## 5. Thermal Conductivity Reduction in Bulk Materials

To reduce the lattice thermal conductivity in bulk thermoelectric materials, two major strategies, alloying [49] and introducing phonon rattlers [57], were proposed in the 1950s and the 1990s respectively. Although their appearances were 40 years apart, both alloy and phonon rattlers essentially share the same idea: introducing atomic disorder either substitutionally or interstitially. In this session, we will focus on the fundamental mechanism, the current understanding and the remaining questions of these two approaches.

**5.1 Alloy**

Alloying is the traditional and probably the easiest way to reduce the lattice thermal conductivity without much degradation to the electrical conductivity. It is well known that the formation of a solid solution between semiconducting or insulating materials usually lead to lower thermal conductivity than the average of the thermal conductivities of the constituent materials in their crystalline form [185]. By substituting lattice sites with different atoms, especially atoms in the same column of the periodic table with similar electronic structures, the impurities can more strongly scatter short wavelength phonons than electrons, resulting in greatly reduced lattice thermal conductivity and well-preserved electronic properties. The alloying effect is particularly large when the mass ratio between the matrix atom and substitutional atom is large. For example, the lattice thermal conductivities of Si and Ge at 300K are rather high (~150 W/mK and ~60 W/mK). The measured thermal conductivity of $Si_{1-x}Ge_x$ can sharply fall down to below ~10 W/mK [48], which led $Si_{1-x}Ge_x$ to become a useful thermoelectric material.

Theoretical studies of phonon transport in alloy systems are limited due to the breakage of the long-range translational symmetry. Thermal conductivites of alloys are typically modeled using the virtual crystal approach introduced by Abeles [186], assuming that the alloy forms a new crystal that is an average of the crystal structures of the constituent materials, which clearly requires them to be of the same crystal type. The new crystal structure has its own corresponding dispersion. The mass disorder and anharmonicity are both treated as first-order perturbations. Thermal conductivity is modeled following Callaway and Klemens' approaches, using Matthiessen's rule to combine different scattering mechanisms. Reduction in thermal conductivity is included in the scattering mechanism. The phonon impurity scattering rate is usually taken into account using the Raleigh scattering model [187] in which the scattering rate is assumed to be proportional to the fourth power of the phonon frequency, with the proportionality constant depending on the mass difference contrast and the atomic volume contrast, or using the improved Tamura model that incorporates the actual phonon density of states [188].

In addition to analytical modeling, numerical attempts have also been carried out in recent years. NEMD simulations based on empirical potentials yielded large discrepancies with measured values for SiGe alloys [189]. Using force constants from DFPT in combination with the virtual crystal model, Garg *et al.* [155] were able to reproduce the experimental thermal conductivity of $Si_{1-x}Ge_x$. The same approach has also been applied to $PbTe_{1-x}Se_x$ [157]. Using force constants from first-principles calculations while varying the atomic masses to account for different half-Heusler alloys, Shiomi *et al.* [74,158] performed EMD simulations, and reached reasonably good agreement with experiments.

Despite the success of the virtual crystal model in matching experimental data, the physical picture is not fully captured by simply averaging the quantities and questions regarding its validity remain. First, Raleigh scattering is only applicable when the wavelength is much longer than the defect dimension [190], which is on the order of atomic sizes for alloys. For high frequency phonons the Raleigh model may not be valid. Second, in the Abeles model [186], atomic mass difference relative to the virtual crystal does not distinguish whether the impurities are heavier or lighter, which does not take into account whether the scatterer frequency is within or above the phonon bands of the matrix. Moreover, it is unclear how the scattering picture transitions from isolated scatterers in the dilute limit to dependent scattering in the non-dilute mixing limit, which is similar to localized and extended states in electronic materials [191,192]. The impurity scattering rates from first-order perturbation deviate from the full perturbation solutions at high frequencies. One also questions why the perturbation theory should hold in the non-dilute limit. Additionally, the proper treatment of phonon group velocities is under debate. Therefore, a thorough understanding of alloy systems is needed.

**5.2 Phonon Rattler**

Slack [57] proposed that skutterudites with open cages can be filled with other atoms which oscillate locally to scatter phonons in the host. Unlike the guest atoms in alloys, which form normal crystal bonds, the guest atoms serving as phonon rattlers are weakly bonded to the oversized cage structure and vibrate locally. Although the rattling idea has led to large progress in ZT of skutterudites, understanding how these extrinsic atoms reduce thermal conductivity has been debated and is still not fully understood.

First, within the phonon rattling picture, there have been different views on whether the localized modes of guest atoms acting alone or the resonant phonon scattering essentially causes the low thermal conductivity. Sales *et al.* [193] suggested that the localized incoherent vibration of guest atoms were responsible for the large reduction in thermal conductivity. The rattler induced localized modes were experimentally observed in filled skutterudites by inelastic neutron scattering [194] and nuclear inelastic scattering [195]. On the other hand, resonant scattering picture was proposed [196], in which only those phonons with frequencies close to the localized rattler frequency would strongly interact with the local mode. Inelastic neutron scattering experiments suggested that fillers of different chemical nature in multi-filled skutterudites provided a broader range of resonant scattering as the quasi-localized modes of fillers were at frequencies sufficiently different from one another [197]. Wang *et al.* [198] conducted ultrafast time-resolved optical measurements to investigate vibrational behaviors of filled antimony skutterudites in time domain. They identified resonant interaction between guest and host atoms as the main cause for the thermal conductivity reduction.

Second, there are additional factors including lattice disorder and point defects, which cannot be decoupled from the rattling effect. Although Slack's idea is that the filler atoms jiggle in the cage at a distinct frequency to scatter phonons of the host, Meisner *et al.* [199] proposed that the reduced thermal conductivity can also be explained by the virtual crystal model. The even stronger reduction in the thermal conductivity of multi-filled skutterudites may also be attributed to the additional point defect scattering from the extra mass fluctuation at the void site [200].

Third, the disordered glass-like picture by rattling modes was questioned by the quasi-harmonic behavior observed by Koza *et al.* [201]. They found coherent coupling between the filler atoms and the host lattice in $LaFe_4Sb_{12}$ and $CeFeb_4Sb_{12}$ using neutron scattering experiments and *ab initio* powder-averaged lattice dynamics calculations. Their finding defied the rattling picture of filler atoms. The coherent low energy phonon modes were thus believed to preserve the phonon crystallinity and contribute to the relatively high thermal conductivity in p-type skutterudites.

In addition, Zebarjadi *et al.* [202] applied MD simulations for the thermal conductivity of filled skutterudites with a simple 2D model. It was found that the decay of the thermal conductivity mainly comes from the flattening of phonon bands, i.e. smaller phonon group velocity, due to the filler atoms.

These controversies, together with the puzzling alloy picture discussed earlier, means that we do not fully understand how impurities in crystals go from individual isolated scattering centers to forming new band structures.

## 6. Thermal Conductivity Reduction in Nanostructured Materials

In this session, we will explore the thermal conductivity reduction in different nanostructures, with the goal of harnessing phonon interface scattering while leaving electronic properties unaffected. Unlike alloys, which scatter very short wavelength phonons, nanostructures can scatter longer wavelength phonons over a wider range at interfaces and surfaces, and effectively suppress the thermal conductivity.

### 6.1 1D Nanostructure: Nanowires

Since the Hicks and Dresselhaus prediction, significant efforts have been devoted to studying thermoelectric effects in different nanowires. Li *et al.* [203] conducted the first thermal conductivity measurement of individual Si nanowires and observed two orders of magnitude reduction from the bulk value. Subsequent experiments on rough Si nanowires [97] reported even lower thermal conductivity, close to the amorphous limit. In addition to Si nanowires, other types of nanowires have also been reported to have lower thermal conductivities [91-96].

The increased surface-to-volume ratio in 1D nanostructures creates much more severe phonon-boundary scattering than in bulk materials. Phonon-boundary scattering can readily account for the thermal conductivity reduction in most nanowires reported in literature. Several theoretical calculations using MD [204-207], BTE solution [208-215] and Landauer formulism [216,217] have been carried out and reached good agreements with experimental data for smooth nanowire with diameters > 20nm. To apply the BTE solution, one can collect the phonon spectral properties by performing lattice dynamics calculations [209], or traditional Monte Carlo (MC) simulation[208], or MC sampling of the phonon free paths [218].

Nevertheless, the physical picture underlying the lower than expected experimental thermal conductivity and the surprising linear temperature dependence of Si nanowires [203] and Ge nanowires [99] with diameters ~ 20 nm and the phonon-roughness scattering effects [97] are thus far not fully understood. Admittedly, the measurements of thin nanowires are very sensitive to non-uniformity of the sample diameters and to surface conditions, which become more difficult to control as diameters decrease. This uncertainty adds to the complexity for theoretical modeling even though the surface roughness effects have recently been quantified experimentally for Si nanowires with diameters~70nm [219].

The theoretical difficulties mainly lie in the proper treatment of phonon-boundary scattering and phonon confinement effects. Molecular dynamics simulations automatically incorporate both phonon confinement and boundary scattering effects, but the diameter that can be simulated is limited. To tackle the phonon confinement effect that causes the phonon spectrum modification, it was suggested to adopt the complete dispersion of nanowires instead of bulk materials [208,211], but only modifying the dispersion from the bulk is not enough [210]. How the surface roughness affects the phonon confinement, in addition to specular or diffuse interface scattering, are open questions. One key challenge is how one can build proper surface structures to best represent the experimental samples. Besides, the accurate modeling of intrinsic phonon lifetimes from first-principles calculations [127,209], the importance of optical phonons in naowires [209], the core defects and the surface oxide layer [206] should also be considered.

## 6.2 2D Nanostructure: Thin-Film Superlattices

Quantum well and superlattices were studied to enhance the Seebeck coefficient via the sharp feature in the electron density of states [30] and to reduce the lattice thermal conductivity. Surprisingly, the thermal conductivity of superlattices was able to be reduced to below the alloy limit [84-89]. Chen [82] argued that the minimum thermal conductivity theory proposed by Slack [125] and Cahill [126] may not apply to low-dimensional materials due to the directionally dependent phonon properties arising from the anisotropy, such as angularly dependent reflectance at interfaces, and provided some initial evidence based on Si/Ge superlattices. Conclusive data was provided by Cahill's group [220,221].

In addition to superlattices, thermal conductivity of individual thin films has also been studied, especially Si films. Goodson and his coworkers [222-226] systematically measured the thermal conductivity of Si thin films down to 74 nm. Their modeling showed that the Fuchs-Sonderheim model on classical size effects can well explain the experimental data, which is quite different from the controversies in Si nanowires. More recently, ultrathin Si films down to 8 nm were used to extract phonon dispersions [227] and phonon lifetimes [228], and to observe quasi-ballistic transport [172].

Theories developed to understand the large thermal conductivity reduction in superlattices can be categorized into an incoherent particle picture and a coherent wave picture. By solving the BTE for partially specular and partially diffuse interface scattering of phonons under the relaxation time approximation, Chen [229,230] concluded that the diffuse and inelastic interface scattering was responsible for the strong thermal conductivity reduction of Si/Ge [84] and GaAs/AlAs [88] superlattices. Ward and Broido [231] applied an exact solution of the inelastic BTE to calculate the intrinsic lattice thermal conductivity of Si/Ge and GaAs/AlAs superlattices by assuming perfect interfaces. They obtained higher thermal conductivities than from the relaxation time approximation approach, implying the rather important effect of interface scattering from imperfections in depressing the thermal conductivity. Another picture is coherent transport. The new periodicity in the superlattice introduces modified phonon dispersions, including phonon miniband formation originating from interference of phonon waves [232], phonon confinement arising from spectra mismatch [233], and group velocity reduction due to phonon reflection [234]. These effects lead to decreased thermal conductivity in the direction perpendicular to the film plane [235], but do not significantly reduce the in-plane thermal conductivity. As period size approaches the extremely small periods, phonon tunneling or simply alloying [235] leads to a recovery of thermal conductivity with decreasing period thickness. Models based on perfect interfaces and phonon dispersion modification, however, generally cannot explain the experimental data observed in both in-plane and cross-plane directions, suggesting the importance of interface roughness. Combining the wave and particle pictures using models [236,237] or MD simulation [238], one can explain the observed trends in different superlattices [84,86,87,89] that the thermal conductivity decreases first as period size increases, reaches a minimum at a few nanometers, and then increases with increasing period size.

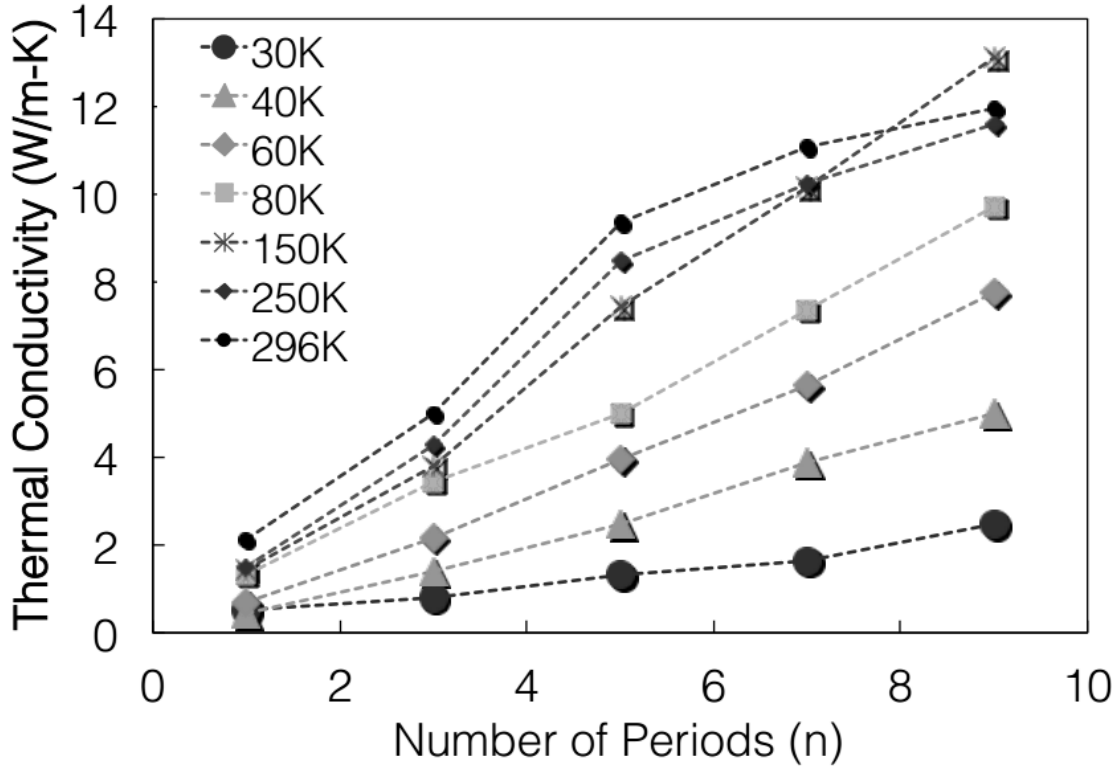

Fig.13 Measured thermal conductivity of GaAs/AlAs SLs as a function of number of periods in the SL for different temperatures. Below 150K, the linearity of the thermal conductivity versus length suggests that phonon heat conduction in these SLs is coherent [239].

Although coherent phonon heat conduction has long been raised in theory and observed for individual phonon modes [240], the experimental demonstration of coherent phonon transport in superlattices has been lacking. Although diffuse scattering at interfaces was responsible for thermal conductivity reduction, past BTE based models still assume a specular component [229], i.e. partially specular and partially diffuse interfaces. A key question is whether one can observe coherent phonon contributions to thermal conductivity. In general, wave effects on thermal conductivity have not been demonstrated clearly, although their importance was speculated in some past experiments in nanowires [203] and nanomesh structures [241]. To answer this question, Luckyanova *et al*. [242] measured thermal conductivities of superlattices by fixing the period length while changing the number of periods. If the interface scattering is completely diffuse, they expect to see a thermal conductivity independent of the number of periods. If transport is coherent, they expected to see that thermal conductivity increases linearly with the total thickness of the superlattice. Figure 13 shows their experimental observation. This experiment, together with lattice dynamics based simulation, shows that low-frequency phonons actually have much longer MFPs than the total thickness of superlattice, and their transport is coherent. If one can destroy the coherence of these low frequency phonons, further reduction in the lattice thermal conductivity is possible.

## 6.3 3D/bulk Nanostructures: Nanocomposites

The studies on the superlattice studies indicate that periodicity is not essential for the thermal conductivity reduction whereas interface scattering creates a more profound decrease [82]. Nanocomposites with a high density of interfaces offer a cost-effective alternative to superlattices as potential thermoelectric materials. Simulation further suggests that the thermal conductivity of nanocomposites can reach below the alloy limit [102,243]. The presence of nanoparticles provides an effective scattering mechanism for mid/long wavelength phonons that contribute heavily to the thermal conductivity.

The theoretical studies on phonon transport in nanocomposites are rather demanding. Kim and Majumdar [244] proposed an approximate analytical solution to estimate the phonon scattering cross section of very small nanoparticles (d=4nm) by bridging the Rayleigh limit and the geometrical scattering limit. They found that large size distribution can more efficiently scatter phonons of a wide range of spectra.

To treat thermal transport in nanocomposites, the first thing one needs to address is phonon transport across interfaces as phonon reflections at the interfaces create interfacial resistance. Thermal boundary resistance depends on the phonon equilibrium density and the phonon transmittance across the interface. The unknown interface condition of the experimental samples adds major complexity for understanding and predicting interfacial transport. Phonon interface transmittance has defied agreements between models and experiments. There have been two different models for the phonon transmittance at an interface: the acoustic mismatch model (AMM) [245] and the diffuse mismatch model (DMM) [246]. As a continuum model, the AMM assumes that phonons undergo specular reflection or transmission at the interface and is valid in the long-wavelength limit. The DMM, on the other hand, assumes not only purely diffuse scattering at the interface, but also equivalence between phonon reflectance from one side to the transmittance from the other. Neither the AMM nor the DMM works well in predicting interface thermal boundary resistances. Using MD at zero temperature [247-252], wave-packets can be created and the phonon transmittance can be obtained by tracking the energy transmitted and reflected after encountering an interface. Although easy to implement, it is computationally expensive since one separate MD simulation is needed for every incoming phonon mode except the multiple phonon wave packet simulations [248]). In addition, it cannot capture a wide angle of incidence, which requires a large lateral size that is difficult to achieve in MD simulations. Linear lattice dynamics calculations [253-256] have been performed to extract the mode-dependent phonon transmittance by solving the reflected and transmitted wave functions according to boundary conditions. However, this method is difficult to implement for complex atomic structures. As an alternative and more computationally efficient approach, a Green's function method dedicated to solve for the response from a point source perturbation is employed to compute the phonon transmission function that can be easily related to transmittance via dividing the transmission function across an interface by its counterpart in the incident material. A general formulation and full derivation have been detailed by Mingo [257,258] and Zhang et al. [259]. Although several other studies are under the same Green's function framework [260-266], only a few studies [263-265] incorporated first-principles force constants into the Green's function calculation. Tian et al. [265] applied the first-principles based Green's function approach to investigate single interfaces between bulk Si and Ge and found that the effects of interface roughness stemming from atomic mixing can even enhance the phonon transmittance due to a smoother transition of the density of states. This conclusion does not conflict with the discussion in the previous section that diffuse scattering reduces thermal conductivity of superlattices, since diffuse scattering destroys the coherence of phonons in superlattices.

The theoretical efforts to access the effective thermal conductivity of nanocomposites mainly fall into three categories: the effective medium approach (EMA), molecular dynamics simulations, and the Boltzmann transport equation.

Nan et al. [267] developed a general effective medium approach (EMA) for the effective thermal conductivity of arbitrary shaped particulate composites as a function of particle size, shape, orientation distribution, volume fraction and thermal boundary resistance. However, this macroscopic model is based on diffusive transport theory that is not always valid at the nanoscale. Minnich and Chen [268] derived a modified EMA formulation for spherical particulate nanocomposites by including the interface density. Their results showed that the interface density dominated the effective thermal conductivity in nanocomposites, where the thermal boundary resistance played a critical role. Ordonez-Miranda et al. [269] extended the modified EMA model of Minnich and Chen [268] for spheroidal inclusions. EMA is convenient to employ, but it requires thermal boundary resistance as an input and it becomes difficult to deal with complicated nanostructures.

Molecular dynamics simulations, on the other hand, allow the direct calculation of thermal boundary resistance across different interfaces [270-276] and the effective thermal conductivity of various nanocomposites [274,277,278]. Using NEMD, the thermal boundary resistance is obtained from the temperature discontinuity at an interface and the effective thermal conductivity is derived from the temperature drop over the sample size. Using EMD, the effective thermal conductivity is given by the well-known Green-Kubo method while the thermal boundary resistance is derived within the framework of the linear response theory [273]. Nevertheless, the thermal boundary resistance determined from MD is an integral quantity which cannot provide the detailed phonon behaviors at interfaces. The empirical potentials adopted in MD simulations also hold it back to certain extent. The small accessible size of the MD method continues to be the main obstacle of their usefulness in complicated composite nanostructures.

Yang et al. [102,215] computed the effective thermal conductivity of periodic two-dimensional (2D) nanocomposite with square silicon nanowires in germanium as well as core-shell and tubular nanowires by numerically solving the BTE under a single mode relaxation time approximation. Prasher [279,280] solved the BTE analytically for simple geometries. For more complicated nanostructures, Monte Carlo (MC) simulation is a useful tool to solve the BTE by tracking the phonon energy bundles as they drift and collide through the computational domain. Jeng et al. [243] used MC simulations to study 3D phonon transport in nanocomposites using a gray-medium approach and showed that the effective thermal conductivity of nanocomposites can reach below the alloy limit [109]. More efficient MC techniques using energy-based variance-reduction were developed to simulate multidimensional phonon transport [281].

The BTE and MC examples described above assume frequency independent MFP and diffuse interface scattering. To accurately model the phonon transport in complex nanostructured materials, the mode-dependent phonon MFPs and interface transmittance from first-principles calculations are desired. Simulation tools already developed, such as extracting the MFPs from first-principles and phonon transmittance from the Green's function method, can in theory provide these input parameters. However, a key challenge is to relate interface conditions to simulations. Past experiments [282] and modeling [283] show that the thermal boundary resistance depends sensitively on imperfections. With the excellent advances on bulk phonon MFPs and encouraging progresses on interface transmittance, both from DFT, multiscale simulation of phonon transport in nanostructures is in sight.

## 7. Heat transfer in thermoelectric devices and systems

Even if thermoelectric materials with high ZT are developed, there are still many device-level challenges to implement thermoelectrics into real applications. In fabricating thermoelectric modules, many technical issues like durability and parasitic losses should be considered. In designing whole systems, heat transfer control should be included to match the heat flux between the heat sink/source and thermoelectric modules. In fact, for many applications, one should consider the device together with the system because system requirements may affect device design. Innovative applications of current materials, although not competitive in prime power and HVAC, are crucial for thermoelectric technology, and arguably, even more urgent than improving ZT. In this section, we discuss heat transfer related issues in thermoelectric modules and systems.

### 7.1 Thermoelectric devices

The standard configuration of thermoelectric devices shown in Fig. 4 is greatly simplified. To connect n-type and p-type pellets together to form a robust device, electrical contacts must be made to the n-type and p-type pellets. Figure 14 illustrates more details at the interface between a thermoelectric pellet and the header. Adjacent to a pellet is typically a diffusion barrier that prevents the electrode material from diffusing into the pellets, which may change their properties, followed by electrodes that join the p-type and n-type legs. The electrodes could also be made of multilayers such that they can bond to the pellets on one side and to an electrically insulating substrate on the other side such that heat can be applied to the pellets, and may contain solders to achieve the bonding. These additional layers introduce electrical and thermal resistances due to their finite thickness and also due to the mismatch of electronic properties at the interfaces, which can significantly reduce the efficiency and COP below that of ideal device expressions in Sec. 2. Due to such resistances, thermoelectric devices should not be made very thin [29]. Min *et. al* [284] further considered thermal/electrical contact resistances in the COP calculation of thin film thermoelectrics, and showed total efficiency can be largely deteriorated by contact resistances. A recent report on direct on-chip cooling by thermoelectrics shows thermal contact resistance is more important than the electrical contact resistance in the useful operation conditions [285].

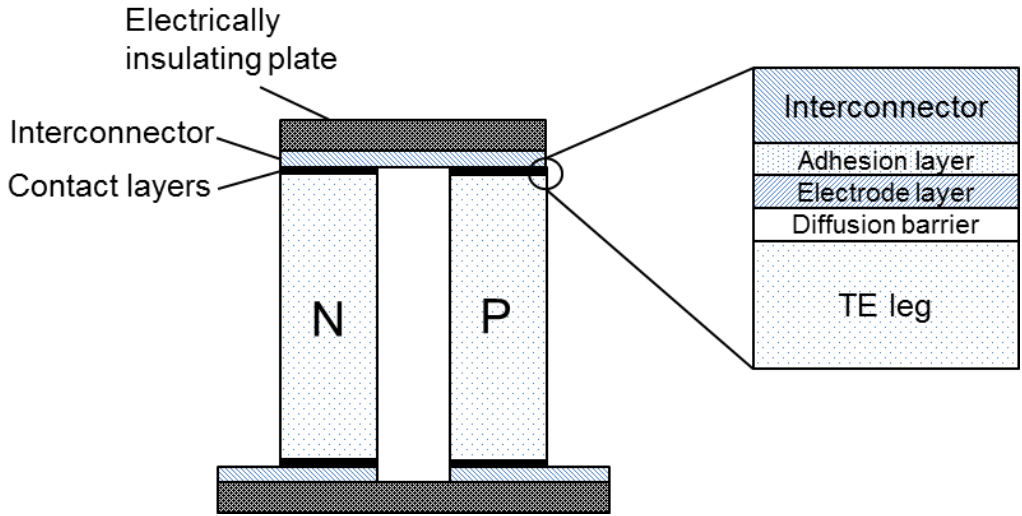

Fig. 14 Typical interface layers between the thermoelectric pellet and the interconnector.

Thermomechanical issues often dictate the choices of electrodes materials and device geometries, especially for power generation applications in which the device experience a large temperature drop. Thermoelectric modules consist of many different components such as thermoelements, interconnectors, and packaging enclosures. Thermal expansion mismatch of the materials causes severe stress. The thermomechanical challenges are especially acute when thermoelectric applications frequently undergo large thermal cycles and mechanical vibrations, such as in the automotive waste heat recovery systems. The large amount of thermal stress can lead to the degradation and even breaking of the contacts between thermoelements and interconnectors, thus resulting in poor performance or failure of the device. As an evidence of the degraded contact, it was reported that total electrical resistance increases with the number of cycles and small cracks form at the interface between thermoelectric pellets and interconnectors [286,287]. To prevent thermal stress within the device, materials with similar thermal expansion coefficients should be chosen. If that is not possible, the thermoelectric module should be designed to minimize those stresses. For example, the T-shunt design in Fig.15(b) may be more durable than a conventional Π-type thermoelectric module under thermal cycles [18]. Interface bonding strength can be improved by proper treatment of the surfaces [288,289].

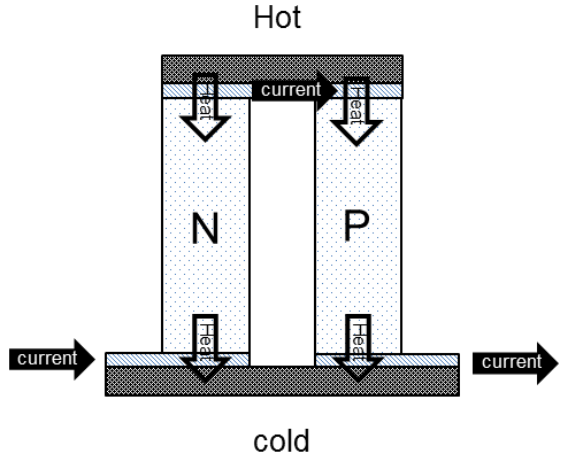

(a)

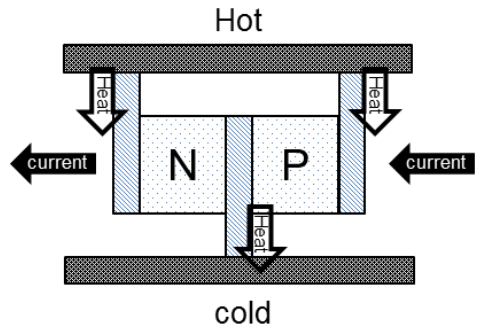

(b)

Fig. 15 Comparison of (a) the conventional Π-type design and (b) the T-shunt thermoelectric design [18]

Interconnectors between p-type and n-type pellets are usually made with metals for high electrical and thermal conductivity, but the metal atoms tend to diffuse into the thermoelectric material at high temperature. The diffused metal can change the carrier density and deteriorate the power factor. Moreover, it can also change the material's structure, which can lead to expansion-induced cracking, and delamination of the contacts. To prevent the diffusion problem, some metal layers can be used as a diffusion barrier. Therefore, contacts should be designed to satisfy both having a high diffusion barrier and a low electrical/thermal contact resistance.

The packaging of thermoelectric modules is also a challenging issue. The reason for the packaging is two-fold: 1) to reduce thermal leakage and 2) to protect thermoelectric modules from harsh environment. Thermal leakage refers to heat transport through void volume in thermoelectric modules, not through thermoelectric pellets which is already included in ZT. Recently, a small filling factor, which is the ratio of a thermoelements area and module area, was emphasized since material cost can be reduced remarkably [290,291]. In the design with a small filling factor, the thermal leakage becomes more problematic. To reduce thermal leakage by radiation, one may try to coat surfaces with low emissive materials. Moreover, the surface coating can prevent sublimation of atoms from the thermoelectric materials. However, for the surface coating to be applied onto thermoelements, it should be ensured not to affect thermoelectric properties. For example, some metals can be used for the surface coating since they have low emissivity. However, they will also shunt electrical and thermal transport if the metal layer is not thin enough. In addition, air conduction can be a major source of thermal leakage for low to intermediate temperature applications. Vacuum packaging may be needed. Regarding the second reason for the packaging, protection of thermoelectric modules from harsh environments is important. High temperature operation of thermoelectric modules can cause oxidation of thermoelectric materials, interconnectors, etc. Packaging adds additional challenges in terms of thermomechanical stress, thermal leakage through the rims of the package, electrical insulation and thermal interfacial resistance between the packaging materials and the thermoelectric modules.

**7.2 System challenges**

Application of thermoelectric energy conversion for heat to electricity generation requires careful system design, including both technical and cost considerations. A key advantage of thermoelectric devices is their solid-state nature, which makes them scalable for small power as well as large power applications. The major obstacle for their deployment is low efficiency. The best reported device level efficiency is up to 12-14 % [292,293]. Because ~60 % of energy usage is wasted as heat, thermoelectrics is often considered as a solution for recovering the waste heat as electricity. However, due to the low efficiency of thermoelectric generators, if one can use thermal insulation to reduce heat leakage and improve the system energy efficiency, thermoelectrics is not suitable for such applications. Thus, when considering waste heat recovery, one should focus on where the heat is really wasted, and insulation is not appropriate.

The economics of thermoelectric power generation depend on the nature of the heat source. When developing a thermoelectric system including heat exchangers, heat flux through the thermoelements should be large enough to maintain the appropriate temperature difference across the thermoelectric pellets while reducing the thickness of the pellets. Reducing the pellet thickness is necessary to keep the cost down. Heat flux across a 1 mm thick thermoelectric pellet with 100 $^{o}$C temperature difference is ~100 kW/m$^2$. This is a large heat flux. One can potentially deliver this heat flux by concentrating heat through heat conduction in the device's hot and cold sides, and using fins and other heat transfer enhancement methods to collect heat to the hot side and reject heat from the cold side. This means that the filling fraction of the thermoelectric pellets will be small. System optimization is crucial, including proper insulation of open spaces between the pellets, thermal interfaces between the module and the heat transfer surfaces, together with electrical isolation.

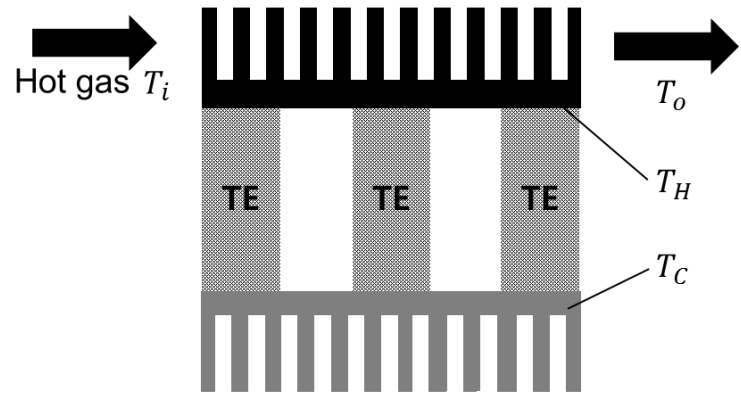

(a)

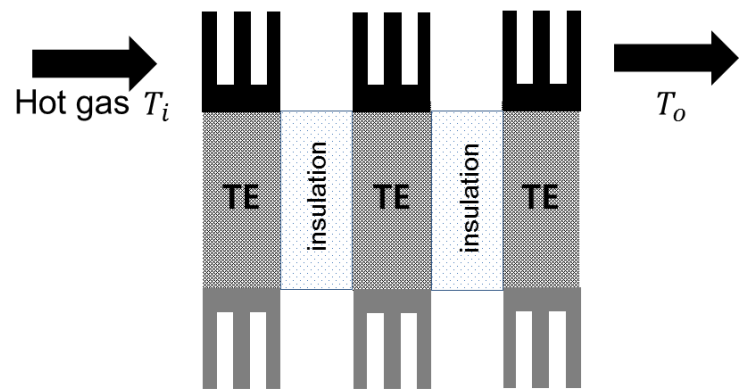

(b)

Fig. 16 Thermoelectric modules with (a) the conventional design and (b) the thermally isolated sub-module design

Another issue in thermoelectric system design is varying heat source/sink temperatures over hot/cold sides of the thermoelectric modules. Taking a typical automobile exhaust waste heat recovery scenario as shown in Fig.16, the heat sink/source temperatures vary along the flow direction due to heat transfer to and out of the module. In one scenario, if the hot side of the thermoelectric module is maintained at a uniform temperature, the exhaust temperature at the exit of the heat recovery section must be higher or equal to the device hot side temperature. The maximum waste heat to electricity conversion efficiency based on the power of the input waste stream is then,

$$\eta = \frac{T_i - T_o}{T_i - T_a} \cdot \frac{T_H - T_C}{T_H} \cdot \frac{\sqrt{1+ZT_M}+1}{\sqrt{1+ZT_M}+T_C/T_H} \qquad (7.1)$$

where $T_i$ and $T_O$ are exhaust temperatures at the inlet and outlet of the thermoelectric power generator, and $T_a$ is the ambient temperature. The first factor represents the efficiency of heat interception from the waste stream and the last two factors are the efficiency of an ideal thermoelectric device operating between $T_H$ and $T_C$. Taking $T_o = T_H$, we obtain the optimal $T_H$ approximately occurs at

$$T_H = \sqrt{T_i T_C} \qquad (7.2)$$

The above rough estimation provides a way to choose the best thermoelectric materials since all the thermoelectric materials have different optimum temperatures and ZT values. This choice, fixes the hot side temperature in between $T_i$ and $T_C$, and hence cannot take advantage of the largest possible temperature difference to generate power. Instead of operating the hot side at one uniform temperature, the thermoelectric module can be divided into many thermally isolated sub-modules, each operating at its own local hot side temperature [294]. In the limiting case of a continuously varying hot side temperature, one can derive differential equations governing the gas stream and hot side temperature and arrives at the maximum efficiency. In Fig.17, we compare the efficiencies obtained for a gas inlet temperature $T_i$=600 °C and device cold side temperature $T_C$=50 °C. The ideal device efficiency operating between 50-600 °C is marked 'Ideal Device 50-600 °C' and shows 15 % when ZT is 1. However, this is not achievable because 1) exhaust gas temperature decreases as it flows and delivers heat to the device, and 2) thermal resistance between the heat source and the thermoelectric module generates additional entropy. If the device operates at a constant hot-side temperature, the optimum temperature is 270 °C and the overall system efficiency is only 5 % when ZT is 1. On the other hand, if we vary the hot side temperature continuously to match the local hot stream temperature, the system efficiency can be raised to 8 % when ZT is 1. The thermally isolated sub-module design can improve the overall efficiency, but there are two major challenges. First, we may need to use different materials along the flow stream since the operating temperature ranges of each thermoelement are different. This probably can be achieved with fine adjustment of carrier concentration. However, if operating temperature ranges are largely different, different types of materials should be used. Second, efficient thermal insulation between thermoelements should be achieved. Otherwise, heat leakage through the insulating layer will decrease the efficiency. Real applications may apply a few stages, each with its own constant hot and cold side temperatures, rather than continuously varying the device temperature.

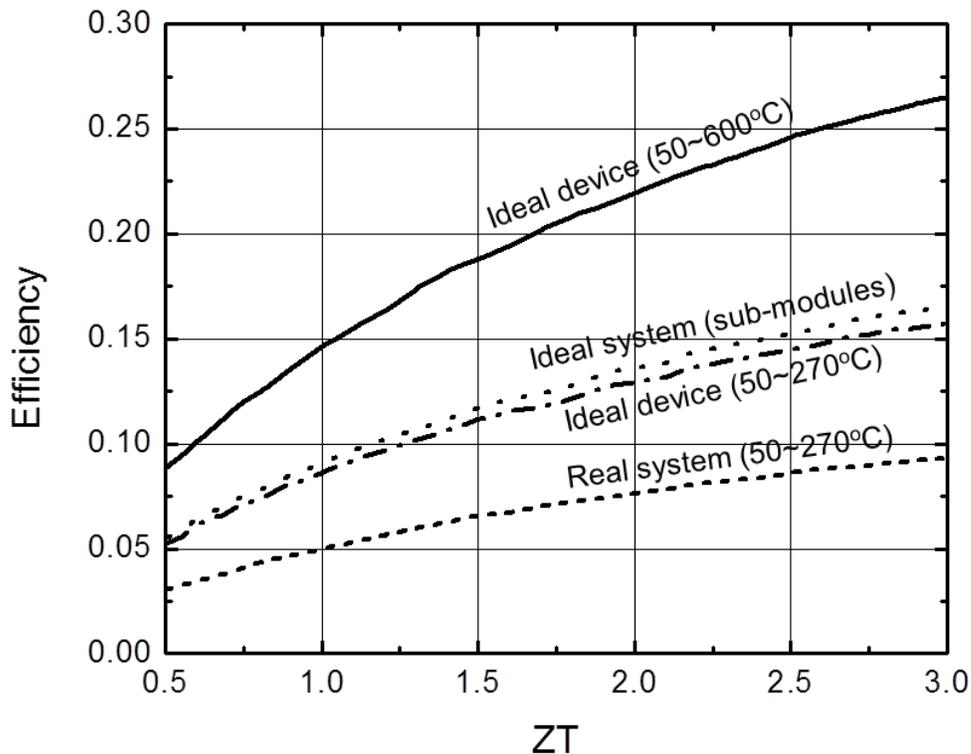

Fig. 17 Efficiency comparison of the conventional module and the thermally isolated sub-modules. Ideal-device is based on heat input to the device, while system is efficiency is based on heat from the incoming waste stream.

Between the thermoelectric modules and the heat transfer surfaces usually lies thermal interface materials. High performance thermal interface materials should have good thermal conductance and high temperature stability, and yet be able to accommodate the thermal stress caused by high temperature operation and large temperature difference. Radio-isotope thermoelectric generators used in NASA space missions employed spring loading between the heat source and the thermoelectric generators [295]. To a certain extent, the space application is simpler since the device operates at a nearly constant temperature difference all the time. Thermal interface materials based on carbon nanotubes have been under development by some groups for such applications [296,297].

Having discussed waste heat applications, we would like to emphasize that thermoelectrics is not limited to waste heat. Thermoelectric energy conversion could also be applied for renewable energy conversion into electricity. In general, thinking about wasted heat availability and entropy generation is probably more appropriate in identifying potential applications. As one example, Kraemer *et al.* recently developed a solar thermoelectric generator employing a small filling factor to make a large heat flux through thermoelements (Fig.18) [291]. Solar insolation is about 1 kW/m$^2$, which is smaller than the preferred heat flux in thermoelectrics by factor of 100-200. By using the small filling factor, they could achieve a large temperature gradient and a respectable efficiency. When such solar-thermoelectric generators are combined with solar hot-water systems as the topping cycle, entropy generation is reduced and economics becomes favorable [298].

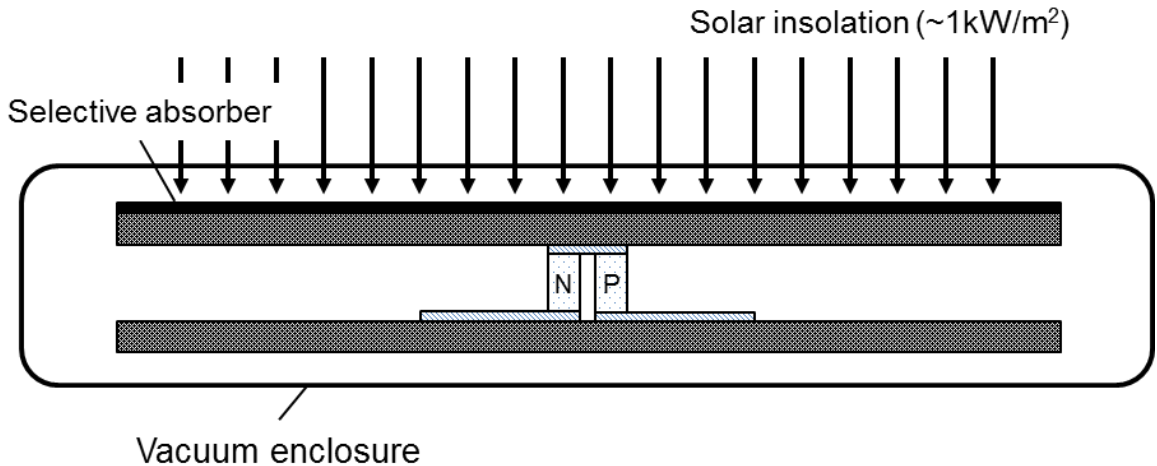

Fig. 18 Schematic picture of the flat-panel solar thermoelectric generator (STEG), which uses thermal concentration, i.e., heat conduction along the absorber plat, rather than optical concentration to achieve a high heat flux.

Regarding to the cost issue, which holds the key to deployment of thermoelectric technology, one should keep in mind that other components in the system like heat exchangers can be a significant portion of the system cost. One interesting feature of thermoelectrics is that output power does not depend on the amount of thermoelectric materials used. For example, a thermoelectric pellet with a small area and short length can generate the same amount of electricity as a thermoelectric pellet with a large area and long length, if their thermal and electrical resistances are the same. Therefore, material cost can be greatly reduced by reducing both filling factor and thermoelement length [290,291]. For example, according to a recent analysis on solar thermoelectric generators, the material cost can be reduced down to $0.03/Wp [298]. Then, the major part of the system cost is due to heat exchangers and other balance-of-plant parts.

We hope that the above discussions illustrate that ample room exists in device fabrication and system innovation. The heat transfer community is well posed to make a difference in this area.

**8. Conclusion**

This review has described the current progresses and challenges surrounding heat transfer in thermoelectric materials and devices, with a special focus on heat transfer physics. One primary strategy to boost ZT is to effectively suppress the lattice thermal conductivity without sacrificing electronic properties because the phononic contribution can be controlled relatively independently without power factor deterioration due to the different MFPs between phonons and electrons. Remarkable progresses have been made to push down the lattice thermal conductivity. In bulk materials, the efforts have focused on alloying and searching for PGEC including using phonon rattlers and complex structures. In nanostructured materials, the increased phonon-boundary or phonon-interface scattering has led to dramatic suppression of thermal conductivity. The excellent progress on first-principle calculations has advanced our fundamental understanding of phonon transport physics in bulk materials, which can be used as guidance to further reduce thermal conductivity. The experimental work on nanostructure materials has moved forward rapidly along with the growing nanostructured material syntheses and measurement techniques.

Despite the excellent progresses, there are still many remaining questions and challenges. One big problem on the experimental side lies in the difficulty and complexity of experimental systems, from microfabricated platforms to optical systems, which prevent most interested groups from entering the fields and limit the reproducibility of experimental data. One major theoretical challenge is the accurate multiscale phonon modeling of nanocomposites since our understanding of phonon-interface scattering is still limited despite encouraging progress on the calculation of mode-dependent phonon interface transmission. Simulation tools for defects and their impact on phonon transport are still lacking. Device fabrication proceeds slowly due to complexity and system-level heat transfer issues. Better and more innovative design based on heat transfer physics to raise the efficiency would enable more practical applications. We hope that this review will kindle broader interest in thermoelectrics within the heat transfer community, promote even faster development of thermoelectric materials and devices and eventually lead thermoelectics to be part of the renewable energy solutions.

## Acknowledgements

This work is supported as part of the S3TEC, an Energy Frontier Research Center funded by the U.S. Department of Energy, Office of Science, Office of Basic Energy Sciences under Award Number DE-FG02-09ER46577 (Z.T. and G.C., for power generation), and OSU MURI under Award Number RF01224242 (S.L., for cooling applications)